\documentclass[a4paper,aps,prd,10pt,twocolumn,nofootinbib,preprintnumbers,
  bibstyle=apsr, longbibliography 
]{revtex4-2}

\usepackage{latexsym,graphicx,amssymb,amsmath,mathrsfs,bbold} 
\usepackage{slashed,mathbbol}
\usepackage[english]{babel}
\usepackage[T1]{fontenc}
\usepackage[utf8]{inputenc} 
\usepackage{color}
\usepackage{bm}
\usepackage{makecell}
\usepackage{booktabs} 
\usepackage{array}    
\usepackage{ragged2e} 
\usepackage{multirow} 
\usepackage{soul}
\usepackage{arydshln}
\usepackage{mathtools}
\usepackage{comment}
\usepackage[svgnames]{xcolor}
\usepackage{cancel}
\usepackage{hyperref} 

\makeatletter
\newcommand{\vast}{\bBigg@{4}}
\newcommand{\Vast}{\bBigg@{5}}
\makeatother

\newcommand{\be}{\begin{equation}}      
\newcommand{\ee}{\end{equation}}      
\newcommand{\bea}{\begin{eqnarray}}      
\newcommand{\eea}{\end{eqnarray}}    
\newcommand{\Tr}{\,\textrm{Tr}\,}
\newcommand{\tr}{\,\textrm{tr}\,}
\newcommand{\Log}{\,\textrm{Log}\,}

\newcommand{\rd}{{\mathrm{d}}}
\newcommand{\ri}{{\mathrm{i}}}

\newcommand{\pvecs}[1]{\vec{#1}\mkern1.75mu'\mkern-1.75mu{}^{2}}
\newcommand{\ppvecs}[1]{\vec{#1}\mkern1.75mu''\mkern-1.75mu{}^{2}}

\long\def\unmarkedfootnote#1{{\long\def\@makefntext##1{##1}\footnotetext{#1}}}
\makeatother

\begin{document} 
\allowdisplaybreaks

\title{Scaling Behaviors in Active Model B+ via the Functional Renormalization Group}
\author{Gergely Fej\H{o}s}
\affiliation{Institute of Physics and Astronomy, E\"otv\"os University, 1117 Budapest, Hungary}
\affiliation{RIKEN Center for Interdisciplinary Theoretical and Mathematical Sciences (iTHEMS), Wako, Saitama 351-0198, Japan}
\author{Zsolt Sz\'ep}
\affiliation{HUN-REN-ELTE Theoretical Physics Research Group, 1117 Budapest, Hungary}
\author{Naoki Yamamoto}
\affiliation{Department of Physics, Keio University, Yokohama 223-8522, Japan}

\begin{abstract}
We study the scaling behaviors of the active model B+ using the functional renormalization group (FRG) approach, based on the nonequilibrium effective action formulated via the Martin-Siggia-Rose path-integral formalism.
We derive the $\beta$ functions for all couplings of the system in generic $d$ dimensions, revealing regulator independence in various contributions to the renormalization group (RG) flow at specific values for $d$.
After identifying specific regions of the parameter space that define submodels closed under RG transformations, we determine all fixed points of potential physical relevance.
We confirm the existence of a bicritical fixed point, which was conjectured within the perturbative momentum-shell RG method for being responsible for the transition from bulk phase separation to microphase separation in active systems.
We argue that, within the FRG approach, global flows significantly differ from those obtained in its perturbative counterpart.
\end{abstract}

\maketitle

\section{Introduction}
In the general philosophy of effective field theories (EFTs), the physics at long timescales and large length scales is governed by the relevant low-energy degrees of freedom and the symmetries, independent of microscopic details. 
From this perspective, a central goal in theoretical physics is to classify universal behavior across various systems based on symmetry principles and EFT data. 
The concept of universality classes is prominently manifested in the theory of critical phenomena in thermal equilibrium. 
For static critical phenomena, scaling laws and critical exponents depend solely on symmetries and spatial dimensionality \cite{goldenfeld2018}. 
In dynamic critical phenomena, the classification introduced by Hohenberg and Halperin \cite{hohenberg1977} remains a cornerstone, organizing systems into various universality classes (e.g., model A, model B, etc.) according to the nature of order parameters and conservation laws.

In contrast, our understanding of universality classes in nonequilibrium systems remains far less complete. 
A particularly novel class of nonequilibrium dynamics arises in active matter \cite{marchetti2013,ramaswamy2017}---systems composed of self-propelled constituents such as cells, bacteria, or artificial microswimmers. 
A defining feature of active matter is the explicit breaking of (generalized) time-reversal symmetry, or Kubo-Martin-Schwinger (KMS) symmetry \cite{kubo1966,martin1959}, which underlies detailed balance in equilibrium dynamics. 
As a result, active systems admit terms in the EFT description that are forbidden in equilibrium, leading to much richer dynamical phenomena.

One of the fundamental nonequilibrium phenomena in active matter is motility-induced phase separation (MIPS) \cite{Tailleur2008,cates2015motility}. 
In the microscopic model of the active Brownian particle (ABP), constituents interact via repulsion, yet can undergo spontaneous phase separation, in which a uniform state can become unstable and separates into coexisting dense and dilute regions. 
While this resembles liquid-gas phase separation in equilibrium systems, it occurs even without explicit attractive interactions. 
The underlying mechanism involves a feedback between local density and motility, whereby particles slow down in crowded environments, effectively inducing an emergent attraction.
The MIPS is considered to terminate at a critical point, characterized by divergent density correlation, on the phase diagram of ABP model as a function of propulsion speed and particle density. 
However, the critical properties near this putative nonequilibrium critical point remain elusive. 
Numerical simulations in Refs.~\cite{partridge2019,maggi2021} report behavior consistent with the equilibrium Ising universality class, whereas other studies~\cite{siebert2018,dittrich2021} observe significant deviations, calling for further clarification.

To capture the universal features of MIPS at the level of continuum EFT, the active model B (AMB) \cite{wittkowski2014,cates2015motility} and later active model B+ (AMB+) \cite{nardini2017,tjhung2018} have been introduced as extensions of model B that incorporate nonequilibrium active terms.
In particular, in Ref.~\cite{tjhung2018} it was reported that in AMB+, a new microphase-separated state can emerge through reverse Ostwald ripening.
A necessary condition for such a state to form is a significant amount of activity in the system.
For small activity parameters, Ostwald ripening is found to operate in a normal fashion, leading to bulk phase separation.
In Ref.~\cite{caballero2018}, a one-loop perturbative renormalization-group (RG) analysis was carried out in an attempt to explain these observations.
It was conjectured that the emergence of a new, nonequilibrium fixed point controls the regions of parameter space where the RG flows either approach the conventional Wilson-Fisher (WF) fixed point or diverge to strong coupling.
The latter may correspond to the aforementioned new state, but the perturbative RG cannot determine whether the flows terminate at a new, strongly coupled fixed point, indicating a second-order transition, or diverge to infinity, signaling a first-order transition. 
Moreover, the very existence of the new nonequilibrium fixed point remains questionable, as the perturbative analysis relies on a simultaneous expansion in both $(d-2)$ and $(4-d)$, which may not yield reliable results when extrapolated to $d=3$.
That is, due to the questionable reliability of perturbation theory in this context, the nature of fixed points and critical phenomena in AMB+ remains an open problem.

In this paper, we study the RG flows of the AMB+ via the nonperturbative functional renormalization group (FRG) approach.
One of the main advantages of this formalism is that it does not require any coupling to be small; consequently, no constraints need to be imposed on the dimensionality. 
Our goal is to determine whether the structure of the coupling flows suggested by perturbation theory persists and to explore the possibility of discovering new fixed points beyond the perturbative regime.

The paper is organized as follows. 
In Sec.~\ref{Sec:basics}, we review the AMB+ and formulate it in terms of a local field theory via the Martin-Siggia-Rose (MSR) formalism. 
Here we also introduce the FRG and discuss how to apply it to the AMB+. 
In Sec.~\ref{Sec:RG}, we review how to obtain flows in the FRG formalism, which is followed in Sec.~\ref{Sec:flow} by the determination of all the $\beta$ functions through diagram technique. 
Section ~\ref{Sec:fixed-points} is devoted to the obtained fixed points, including those belonging to submodels of AMB+ that close under the RG.
In that section, we also compare our results with the perturbative momentum-shell RG. 
Conclusions can be found in Sec.~\ref{Sec:concl}, while calculational details of the fluctuation integrals are presented in the Appendix.

\section{Basics}
\label{Sec:basics}

The goal of this section is threefold.
First, we introduce the AMB+ and then discuss how to formulate it into a local field theory via the MSR formalism, and finally, we review the FRG technique.

\subsection{AMB+}
\label{Sec:AMB+}
In this paper we consider the AMB+, which describes the dynamics of a conserved scalar order parameter (density) field $\phi(t,\vec{x})$ through the corresponding continuity equation,
\be
\label{Eq:cont}
\partial_t \phi = -\vec\nabla\cdot \vec{J}\,,
\ee
where $\vec{J}$ is the conjugate current to $\phi$, 
\be
\label{Eq:current}
\vec{J} = -M\vec\nabla \mu + M\zeta \big(\Delta\phi\big)\vec\nabla\phi + \vec\xi\,.
\ee
Here $M$ is the mobility, which is assumed to be constant.
The current is realized, on the one hand, through spatial dependence of the chemical potential $\mu$ (first term) and, on the other hand, through some kind of activity (second term), which is a genuinely nonequilibrium contribution that cannot be written as a gradient of a local function. 
The current also includes a random force, modeled through a Gaussian noise, $\vec{\xi}$, with zero mean and correlation
\be
\label{Eq:noise}
\langle\xi_i(t,\vec{x})\xi_j(t',\vec{x}')\rangle = C\delta_{i j}\delta(t-t')\delta(\vec{x}-\vec{x}')\,.
\ee

It is important to stress that $\mu$ is a sum of its passive and active counterparts, 
\be
\label{Eq:mu}
\mu=\mu_{\rm P}+\mu_{\rm A}\,,
\ee
where $\mu_{\rm P}$ is the chemical potential of model B, which can be written as the functional derivative of a Landau-Ginzburg type free energy,
\be
\label{Eq:free_energy}
\mathcal{F}[\phi] = \int \rd^d x \bigg[\frac{a}{2}\phi^2+\frac{u}{4}\phi^4+\frac{K}{2}\big(\vec\nabla\phi\big)^2\bigg]\,,
\ee
as
\be
\label{Eq:muP}
\mu_{\rm P} := \frac{\delta \mathcal{F}[\phi] }{\delta \phi} = a\phi+u\phi^3 -K\Delta\phi\,.
\ee
On the other hand, 
\be
\mu_{\rm A} := \lambda\big(\vec\nabla\phi\big)^2 + \frac{\nu}{2}\Delta (\phi^2)
\ee
cannot be expressed as a functional derivative and hence represents an active contribution.

Notice that the activity-induced nonlinearities in the chemical potential comes about in two distinct terms and couplings ($\lambda$ and $\nu$), making the AMB+ contain three sources of activity ($\zeta, \lambda$, and $\nu$). 
The term associated with the $\nu$ parameter can also be interpreted as the leading density correction to the $K$ coefficient, which was shown to be insignificant at the mean-field level \cite{tjhung2018}. 
As we will see shortly, in a RG treatment, this term must be included to ensure that the flows properly close under scale transformations (see also Ref.~\cite{caballero2018}).

The model described by Eqs.~\eqref{Eq:cont}--\eqref{Eq:mu} is symmetric under the transformation $(\phi, \lambda, \nu, \zeta) \to -(\phi, \lambda, \nu, \zeta)$. 
The terms proportional to the active couplings, that is, $\lambda$, $\nu$, and $\zeta$, break generalized time-reversal symmetry, hence, they violate detailed balance [see Eq.~(\ref{Eq:TRS})].
Their contribution to $\vec{J}$ is of the same order in the field and gradients and we note that there are no other independent terms that could introduce activity into the system.
Here and below, we assume that only these active terms break detailed balance. This means that by setting $\mu_{\rm A}=\zeta=0$, one recovers model B, which satisfies $C=2 M T$, with $T$ being the temperature due to the fluctuation-dissipation (FD) theorem. On the other hand, AMB is obtained when $\zeta = 0$. Note that the first study on AMB in Ref.~\cite{wittkowski2014} considered $\zeta=\nu=0$ when describing phase separation dynamics.

\subsection{MSR formalism}

In this subsection, we review the MSR path-integral formalism \cite{martin1973,janssen1976,deDominicis1978} (see also Ref.~\cite{tauber2014}), which formulates the AMB+ in terms of a path integral belonging to a local field theory.
\footnote{More recently, a fully systematic derivation of the similar nonequilibrium EFT was developed in Ref.~\cite{crossley2015}, in which the KMS symmetry plays an essential role. See also Refs.~\cite{landry2023, Huang:2023eyz} on the construction of EFTs for active matter based on the KMS symmetry.}
Our main motivation for this construction is that, through this formalism, we can apply the FRG technique to the system in a straightforward fashion. 

Considering an observable ${\cal O}$, let us calculate its mean value due to our nonequilibrium stochastic dynamics:
\be
\langle {\cal O} \rangle \propto \int\!{\cal D}\vec{\xi}\,{\cal O}[\phi_{\rm sol}]\, e^{-\frac12\int_{\vec{x},t} \xi_i C^{-1} \delta_{ij} \xi_j}\,.
\ee
Here we omitted the normalization factor, and we stress that the observable ${\cal O}$  needs to be considered at the solution of the stochastic dynamics, in principle as a function of the noise. 
By inserting a functional delta function\footnote{Here we entirely drop the Jacobian; see the explanation at the end of this subsection.} and interchanging the integrals, we arrive at
\bea
\label{Eq:O}
\langle {\cal O} \rangle &\!\propto\!& \int\!{\cal D}\vec{\xi}\; {\cal O}[\phi_{\rm sol}] e^{-\frac12\int_{\vec{x},t}\!\xi_i C^{-1} \delta_{ij} \xi_j}\! \int\!{\cal D}\phi\, \delta(\partial_t \phi + \vec{\nabla}\cdot \vec{J}\,) \nonumber\\ 
&\!=\!&\int\!{\cal D}\phi \!\int\!{\cal D}\vec{\xi}\; {\cal O}[\phi] \,\delta(\partial_t \phi + \vec{\nabla}\cdot \vec{J}\,)\; e^{-\frac12\int_{\vec{x},t}\!C^{-1} \vec{\xi}\,^2}.
\eea
The main point is that now the ${\cal O}$ observable can be evaluated at $\phi$, instead of $\phi_{\rm sol}$, since the delta function ensures that the integrand is nonzero only for $\phi = \phi_{\rm sol}$. 
Rewriting the delta function using its functional Fourier representation,
\be
\label{Eq:delta}
\delta(\partial_t \phi + \vec{\nabla}\cdot \vec{J}\,) = \int {\cal D} \pi \, e^{\ri\int_{\vec{x},t} \pi (\partial_t \phi + \vec{\nabla} \cdot\vec{J}\,)}\,,
\ee
and substituting Eq.~(\ref{Eq:current}) into Eq.~(\ref{Eq:O}), we arrive at
\bea
\label{Eq:O2}
\langle {\cal O} \rangle &\propto& \int{\cal D}\phi \int\! {\cal D}\pi\,{\cal O}[\phi]\,e^{\int_{\vec{x},t} \pi [\partial_t \phi + M \vec{\nabla}\cdot(-\vec\nabla \mu + \zeta (\Delta\phi)\vec\nabla\phi)]}\nonumber\\
&&\times\int {\cal D}\vec{\xi}\,e^{-\int_{\vec{x},t} (\frac12 C^{-1} \vec{\xi}\,^2-\pi \vec{\nabla}\cdot\vec{\xi}\,)}\,,
\eea
where we performed a $\pi \rightarrow \pi/\ri$ rescaling. 
From now on, we shall refer to $\pi(t,\vec{x})$ as the \textit{response} field.

The $\vec{\xi}$ integral can be done:
\be
\label{Eq:xiint}
\int {\cal D}\vec{\xi}\,e^{-\int_{\vec{x},t} (\frac12 C^{-1} \vec{\xi}\,^2-\pi \vec{\nabla}\cdot\vec{\xi}\,)} \propto e^{\int_{\vec{x},t} \frac{C}{2}(\vec{\nabla} \pi)^2}\,,
\ee
and now we can associate a ${\cal P}[\phi,\pi]$ probability density to each field configuration and calculate averages as
\be
\langle {\cal O} \rangle = \int {\cal D}\phi \int {\cal D}\pi\, {\cal O}[\phi] {\cal P}[\phi,\pi]\,,
\ee
with ${\cal P}[\phi,\pi] \propto {\rm exp}\big(-\int_{\vec{x},t} {\cal L}[\phi,\pi]\big)$, where
\be
\label{Eq:L}
{\cal L} = - \pi \partial_t \phi + M\pi \Delta \mu - M\zeta \pi 
\vec{\nabla} \cdot \Big[(\Delta \phi) \vec{\nabla} \phi \Big] + \frac{C}{2} \pi \Delta \pi
\ee
can be regarded as a Lagrangian density, with $C=2MT$, due to the FD theorem in model B without activity. 

The corresponding action, in the absence of the active terms $\mu_{\rm A}$ and $\zeta$, possesses the generalized time-reversal symmetry (or KMS symmetry):
\be
\label{Eq:TRS}
t \rightarrow - t, \quad 
\phi \rightarrow \phi, \quad
\pi \rightarrow - \pi - \frac{\mu_{\rm P}}{T}\,.
\ee
Here, we used the fact that the term containing the $\phi$-dependent $\mu_{\rm P}$, generated by the shifts in the first term of the Lagrangian, can be written at the level of the action as a total time derivative due to Eq.~\eqref{Eq:muP} and thus can be dropped as a surface term.
The active terms, however, explicitly break the symmetry (\ref{Eq:TRS}).
In other words, they are characterized by the violation of the generalized time-reversal symmetry, as stated at the end of Sec.~\ref{Sec:AMB+}.

After performing a few partial integrations at the level of the action and using Eqs.~(\ref{Eq:mu}) and (\ref{Eq:muP}) in the Lagrangian (\ref{Eq:L}), we get
\begin{align}
\label{Eq:Lag}
{\cal L} &= \phi \, \partial_t \pi - M T (\vec{\nabla}\pi)^2 + M \Delta \pi \Big[-K\Delta \phi + a \phi + u\phi^3 \nonumber\\
&+ \lambda (\vec{\nabla}\phi)^2 + \frac{\nu}{2}\Delta (\phi^2)\Big] 
+ M\zeta (\Delta\phi) \vec{\nabla}\pi\cdot \vec{\nabla}\phi.
\end{align}
On this form of the Lagrangian, one can easily notice that for any constant parameter $c$, the transformation $\pi \rightarrow \pi + c$ does not change ${\cal L}$, which can be interpreted as a global shift symmetry of the system. 
This property will play an important role in the construction of the FRG regulator [see Eq.~\eqref{Eq:reg}].

One can eliminate the mobility by rescaling the time, and the temperature by rescaling the fields and some of the couplings.
On the other hand, unlike in static systems, the parameter $K$ cannot be scaled away in the action $\int_{\vec{x},t} {\cal L}[\phi,\pi]$.
Choosing to eliminate it from the passive part of the chemical potential, we perform the following rescalings:
\be
\label{Eq:rescalings}
\begin{gathered}  
t \to \frac{t}{M}\,,\quad \phi\to\phi\sqrt{\frac{T}{K}}\,,\quad \pi\to\frac{\pi}{\sqrt{T K}}\,,\\
a\to K a\,,\ \ u\to \frac{K^2}{T}u\,,\ \ \big(\lambda,\nu,\zeta\big)\to\sqrt{\frac{ K^3}{T}}\big(\lambda,\nu,\zeta)\,.
\end{gathered}
\ee
Then the Lagrangian takes the form
\begin{align}
\label{Eq:Lag_rescaled}
{\cal L} &= K^{-1}\phi \, \partial_t \pi - K^{-1}(\vec{\nabla}\pi)^2 +\Delta \pi \Big[-\Delta \phi + a \phi + u\phi^3 \nonumber\\
&+ \lambda (\vec{\nabla}\phi)^2 + \frac{\nu}{2}\Delta (\phi^2)\Big] 
+ \zeta (\Delta\phi) \vec{\nabla}\pi\cdot \vec{\nabla}\phi\,.
\end{align}


Finally, we comment on why the Jacobian corresponding to the functional delta function in Eq.~(\ref{Eq:O}) was not taken into account. 
It can be shown that this factor exactly eliminates all one-loop diagrams with closed response loops in a diagrammatic expansion of the effective action (in these diagrams the $\pi$-$\phi$ propagator runs in a closed loop) \cite{tauber2014}. 
That is, we can completely forget about the Jacobian provided that we also drop these diagrams.

\subsection{FRG technique}

The model, defined via Eq.~(\ref{Eq:Lag_rescaled}), as a Euclidean statistical field theory, leads to the following generator functional:
\be
Z[J_{\phi},J_{\pi}] = \int\!{\cal D}\pi {\cal D}\phi\, \exp\Big(-\int_{\vec{x},t}\!({\cal L}+J_\phi \phi+J_\pi \pi)\Big)\,,
\ee
where $J_{\phi}$ and $J_{\pi}$ are source fields conjugate to $\phi$ and $\pi$, respectively. 
The corresponding 1PI effective action will be denoted by $\Gamma[\phi,\pi]$.%
\footnote{Note that we use the same notations for these variables as for the fluctuating fields appearing in the Lagrangian.}
In the FRG formalism the effective action also depends on an external momentum variable, $k$, intended to realize scale separation such that, by definition, $\Gamma_k$ only incorporates fluctuations with momentum larger than $k$. 
This is achieved through a momentum-dependent mass term, usually called the regulator, which is a quadratic form in the space of the dynamical variables (i.e., in our model this is the $\phi$-$\pi$ two-dimensional space), giving large mass to the infrared (IR) modes effectively freezing them, while leaving the ultraviolet (UV) ones untouched. 
For details of the formalism, the reader is referred to the following reviews \cite{Berges:2000ew,Dupuis:2020fhh}. 

In this study, we are interested in whether the effective action, $\Gamma_k$, can show scaling behaviors with respect to $k$. It obeys the Wetterich equation \cite{Wetterich:1992yh}:
\be
\label{Eq:flow}
k\partial_k \Gamma_k = \frac12k\tilde{\partial}_k \Tr \Log \big(\Gamma_k^{(2)}+{\cal R}_k\big)\,,
\ee
where $\Gamma_k^{(2)}$ is the second derivative of $\Gamma_k$ (sometimes called the proper two-point vertex function) and ${\cal R}_k$ is the already-mentioned regulator function, being a $2\times 2$ matrix in the $\phi$-$\pi$ space. 
Note that the trace has to be taken both in the functional and in the matrix sense, and $\tilde{\partial}_k$ acts by definition only on ${\cal R}_k$.

The second derivative matrix, $\Gamma_k^{(2)}$, needs to be calculated for general inhomogeneous background fields of $\pi$ and $\phi$.
The reason why we cannot assume the fields to be homogeneous is that those couplings, which belong to operators containing derivatives, would immediately vanish in the flow equation. 
First we split $\Gamma_k^{(2)}$ into two parts: 
\bea
\Gamma_k^{(2)} = \Gamma_k^{(2)0} + {\cal P}_k,
\eea\\
where $\Gamma_k^{(2)0}$ is the two-point function without interactions, i.e., the part coming from terms in Eq.~(\ref{Eq:Lag_rescaled}) that are quadratic in the field. 
The remaining contributions are defined into ${\cal P}_k$.
The reason for this splitting is that if we now introduce the notation $\Gamma_{k,R}^{(2)0} := \Gamma_{k}^{(2)0} + {\cal R}_k$, then Eq.~(\ref{Eq:flow}) can be rewritten as 
\begin{align}
k\partial_k \Gamma_k &= \frac12 k\tilde{\partial}_k \Tr \Log \big( \Gamma_{k,R}^{(2)0} + {\cal P}_k\big)\nonumber\\
\textcolor{white}{\vbox{\hbox{$\frac{a}{a}$}\hbox{b}}}
&= \frac12 k\tilde{\partial}_k \Tr \Log \big(\Gamma_{k,R}^{(2)0}\big) \nonumber\\
&\quad + \frac12 k\tilde{\partial}_k \Tr \Log \Big( {\mathbb{1}} + \big(\Gamma_{k,R}^{(2)0}\big)^{-1}{\cal P}_k\Big)\,,
\end{align}
where the first term is a field-independent irrelevant constant that can be entirely dropped from the description.
Therefore, we can always work safely with the following equation:
\be
\label{Eq:flow2}
k\partial_k \Gamma_k = \frac12 k\tilde{\partial}_k \Tr \Log \Big( {\mathbb{1}} + \big(\Gamma_{k,R}^{(2)0}\big)^{-1}{\cal P}_k\Big)\,,
\ee
which is much more convenient than the original form (\ref{Eq:flow}), since the inverse of $\Gamma_{k,R}^{(2)0}$ can always be easily calculated, allowing the right-hand side of Eq.~(\ref{Eq:flow2}) to be expressed as the functional Mercator series of the logarithm.
Since $\Gamma_{k,R}^{(2)0}$ is field independent, it can always be assumed to be translationally invariant, which means that a delta function can be factored: $\big(\Gamma_{k,R}^{(2)0}\big)^{-1}(q,p) := (2\pi)^d\delta(q+p){\cal D}_{k,R}(q)$.
Here, $q=(\omega,\vec{q})$ is a $(d+1)$-dimensional variable, including frequency and spatial momentum, which are conjugate variables to time and space, respectively.
Using these notations, the right-hand side of Eq.~(\ref{Eq:flow2}) can be written explicitly as 
\vspace*{-0.5cm}
\begin{widetext}
\begin{align}
\label{Eq:flow3}
\!\!\!\!\!\!\!k\partial_k \Gamma_k &= \frac12 \int_{q_1}\! k\tilde{\partial}_k \tr \Big[ {\cal D}_{k,R}(q_1){\cal P}_k(q_1,-q_1)\Big]\nonumber\\
& \quad -\frac14 \int_{q_1,q_2}\! k\tilde{\partial}_k \tr \Big[ {\cal D}_{k,R}(q_1){\cal P}_k(q_1,-q_2){\cal D}_{k,R}(q_2){\cal P}_k(q_2,-q_1)\Big]\nonumber\\
& \quad +\frac16 \int_{q_1,q_2,q_3}\! k\tilde{\partial}_k \tr \Big[{\cal D}_{k,R}(q_1){\cal P}_k(q_1,-q_2){\cal D}_{k,R}(q_2){\cal P}_k(q_2,-q_3){\cal D}_{k,R}(q_3){\cal P}_k(q_3,-q_1)\Big]\nonumber\\
& \quad -\frac18 \int_{q_1,q_2,q_3,q_4}\! k\tilde{\partial}_k \tr \Big[{\cal D}_{k,R}(q_1){\cal P}_k(q_1,-q_2){\cal D}_{k,R}(q_2){\cal P}_k(q_2,-q_3){\cal D}_{k,R}(q_3){\cal P}_k(q_3,-q_4){\cal D}_{k,R}(q_4){\cal P}_k(q_4,-q_1)\Big]\nonumber\quad\\
& \quad +...\, ,
\end{align}
\end{widetext}
where the remaining trace operations need to be taken in the matrix sense only, and a momentum integral is defined as $\int_{q_i} := \int \frac{\rd\omega_i}{2\pi}\int \frac{\rd^d q_i}{(2\pi)^d}$.
These series formally generate one-loop diagrams, namely, tadpole, bubble, triangle, and box diagrams, which are made of regularized propagators ${\cal D}_{k,R}$ and vertices that are contained in ${\cal P}_k$. 
In the forthcoming discussion, we will only make use of the first four terms, written explicitly in Eq.~(\ref{Eq:flow3}).

At this point, we stress once again that due to the construction of the MSR Lagrangian, all diagrams that are one-loop and made by a single $\pi$-$\phi$ propagator (the so-called closed response loops) need not be taken into account. 
These can only arise from the first term in the right-hand side of Eq.~(\ref{Eq:flow3}), which contains one propagator only.

\section{RG flows in the AMB+}
\label{Sec:RG}

So far our analysis was very general, we have not used our specific model defined in Eq.~(\ref{Eq:Lag_rescaled}). 
At this point, however, we must choose an ansatz for $\Gamma_k$ that is specifically tailored to our theory, as the flow equation does not admit an exact solution in any of its forms. 
In this study, we choose $\Gamma_k$ to be formally equivalent to Eq.~$(\ref{Eq:Lag_rescaled})$, but with scale-dependent couplings, and we also rescale the fields according to $\phi \rightarrow Z^{1/2}_{\phi,k} \phi$ and  $\pi \rightarrow Z^{1/2}_{\pi,k} \pi$.
That is, what we are after is a set of flow equations for a number of six couplings ($K_k, a_k, u_k, \lambda_k, \nu_k, \zeta_k)$ and the aforementioned two wave-function renormalization factors ($Z_{\phi,k}, Z_{\pi,k}$). 
In this study, we will not consider the dynamical critical exponent $z$, and therefore, no separate rescaling factor will be introduced corresponding to time.

In Fourier representation the following momentum-dependent cubic vertex function emerges from the $\sim \pi\phi^2$ terms of the Lagrangian:
\be
\label{Eq:vertex}
\begin{aligned}
{\cal V}_k(q_1,q_2,q_3)=&\big[2\lambda_k \vec{q}_2\cdot \vec{q}_3 + \nu_k(\vec{q}_2+\vec{q}_3)^2\big]\vec{q}_1^{\,2}\\
&-\zeta_k\big[\vec{q}_2\cdot \vec{q}_3(\vec{q}_2^{\,2}+\vec{q}_3^{\,2})+2\vec{q}_2^{\,2}\vec{q}_3^{\,2}\big]\,.
\end{aligned}
\ee
Its first momentum $q_1$ belongs to the $\pi$ field, it is symmetric under the exchange of its last two momenta, $q_2$ and $q_3$, and the sum of momenta is zero.
Then, from Eq.~(\ref{Eq:Lag_rescaled}) we get our ansatz in the form
\bea
\label{Eq:ansatz}
&&\Gamma_k = \Big(\prod_i\int_{q_i}\Big) \bigg[ -Z_{\pi,k}K_k^{-1} \vec{q}_1^{\,2}\; \pi(q_1) \pi(q_2)\bar{\delta}(q_1+q_2)\nonumber\\
  &&\ \ -Z^\frac{1}{2}_{\pi,k}Z^\frac{1}{2}_{\phi,k}(-\ri \omega_1 K_k^{-1}+\vec{q}_1^{\,4}+a_k \vec{q}_1^{\, 2})\pi(q_1)\phi(q_2)\bar{\delta}(q_1+q_2)\nonumber\\
  &&\ \ +\frac12 Z^\frac{1}{2}_{\pi,k}Z_{\phi,k} {\cal V}_k(q_1,q_2,q_3)\;\pi(q_1)\phi(q_2)\phi(q_3)\bar{\delta}\Big(\sum_{j=1}^3 q_j\Big)\nonumber\\
  &&\ \ -Z^\frac{1}{2}_{\pi,k}Z^\frac{3}{2}_{\phi,k}u_k\vec{q}_1^{\,2}\; \pi(q_1)\phi(q_2)\phi(q_3)\phi(q_4)\bar{\delta}\Big(\sum_{j=1}^4 q_j\Big)
  \bigg]\,,
\eea
where we introduced the shorthand notation $\bar{\delta}(q)=(2\pi)^d \delta(q)$. 
Note that $\vec{q}_i$ is a $d$-dimensional momentum conjugate to the $d$-dimensional real space coordinate, while $q_i=(\omega_i, \vec{q}_i)$ is a $(d+1)$-dimensional momentum.
Scale dependence is obtained by projecting the right-hand side of Eq.~(\ref{Eq:flow}) onto all terms that appear in Eq.~(\ref{Eq:ansatz}), matching the coefficients with the left-hand side. 
The six couplings and the two wave function renormalization factors would require eight independent terms in the Lagrangian, whereas we only have seven. This implies that our set of flow equations is underdetermined.\footnote{The $\sim \phi\partial_t \pi$ term cannot be used, as this projection is supposed to determine the dynamical critical exponent, which we do not consider here.} 
However, one notices that, in order for the anomalous dimensions appearing in the RG flows of the couplings in the AMB+ to reduce to those of the passive model when no activity is present, one must choose $Z_{\pi,k}= Z_{\phi,k} =Z_k$. 
From this point onwards, this choice is adopted.

Using Eq.~(\ref{Eq:ansatz}), the free two-point function in Fourier space reads
\be
\label{Eq:gamma20}
\begin{aligned}
\Gamma^{(2)0}_{k,\pi\pi}(q_1,q_2)&=-2 Z_{k}K_k^{-1} \vec{q}^{\,2}_1\bar{\delta}(q_1+q_2),\\
\Gamma^{(2)0}_{k,\phi\phi}(q_1,q_2)&=0,\\
\Gamma^{(2)0}_{k,\pi\phi}(q_1,q_2)&=Z_{k}(- \vec{q}^{\,4}_1-a_k \vec{q}^{\,2}_1+\ri\omega_1 K_k^{-1})\bar{\delta}(q_1+q_2),\\
\Gamma^{(2)0}_{k,\phi\pi}(q_1,q_2)&=Z_{k}(- \vec{q}^{\,4}_1-a_k \vec{q}^{\,2}_1-\ri\omega_1 K_k^{-1})\bar{\delta}(q_1+q_2).
\end{aligned}
\ee
The interaction part, in turn, is obtained as
\be
\label{Eq:P}
\begin{aligned}
\!\!{\cal P}_k^{\pi\pi}(q_1,q_2)&=0,\\
\!\!{\cal P}_k^{\phi\phi}(q_1,q_2)&=-6 Z_k u_k\!\int_{q_3,q_4}\!\!\!\!\vec{q}_4^{\,2} \phi(q_3)\pi(q_4)\bar{\delta}\Big(\sum\limits_{i=1}^4 q_i\Big)\\
&\quad + Z^\frac{3}{2}_{k}\!\int_{q_3}\!\! {\cal V}_k(q_3,q_1,q_2) \pi(q_3)\bar{\delta}\Big(\sum\limits_{i=1}^3 q_i\Big),\\
\!\!{\cal P}_k^{\pi\phi}(q_1,q_2) &=-3 Z_k u_k \vec{q}_1^{\,2}\!\int_{q_3,q_4}\!\!\!\!\phi(q_3)\phi(q_4)\bar{\delta}\Big(\sum\limits_{i=1}^4 q_i\Big)\\
&\quad + Z^\frac{3}{2}_{k}\!\int_{q_3}\!\! {\cal V}_k(q_1,q_2,q_3) \phi(q_3)\bar{\delta}\Big(\sum\limits_{i=1}^3 q_i\Big),\\
\!\!{\cal P}_k^{\phi\pi}(q_1,q_2)&={\cal P}_k^{\pi\phi}(q_2,q_1).
\end{aligned}
\ee

We still need to specify the regulator matrix, for which, in principle, there are various choices. 
Note, however, that, as discussed earlier, the original Lagrangian (\ref{Eq:Lag}) has the global shift symmetry $\pi \rightarrow \pi + c$, which we wish to preserve throughout the RG flow. 
If we broke this symmetry with the regulator, then the RG would generate various new terms not present in the original Lagrangian, a result which we must prohibit.
In order to prevent the violation of the shift symmetry, the IR regulator term must not depend on $\pi$ itself, only on its derivative(s). 
Therefore, we choose only to regulate the $\pi$-$\phi$ component of the two-point function, i.e., the $2\times 2$ regulator matrix, ${\cal R}_k$, is chosen to be off-diagonal leading to the following term in Fourier space
\be
\label{Eq:reg}
\begin{split}
&\int_{q_1,q_2}  \pi(q_1)\phi(q_2) ({\cal R}_{k})_{\pi\phi}(q_1,q_2):= \\
&\int_{q_1,q_2}  \pi(q_1)\phi(q_2) \Big(-Z_{k}\vec{q}^{\,2}_1 R^{\rm L}_k(\vec{q}_1)\bar{\delta} (q_1+q_2)\Big)\,, 
\end{split}
\ee
which is added to the Lagrangian for IR mode suppression.
Here, $R^{\rm L}_k(\vec{q})=(k^2- \vec{q}^{\,2})\Theta(k^2- \vec{q}^{\,2})$ is Litim's regulator function \cite{Litim:2001up}. 
We will see explicitly that this choice properly regularizes the effective action, ensuring the absence of IR divergences.

By inverting $\Gamma_{k,R}^{(2)0}$ using the components in Eq.~\eqref{Eq:gamma20} and the regulator in Eq.~\eqref{Eq:reg}, the regulated version of the two-point function obtained after factoring out a delta function has the following form:
\be
\label{Eq:gamma20inv}
   {\cal D}_{k,R}(q)= 
   \left(\begin{matrix} 0 & -Z^{-1}_k d^{\pi\phi}_{k,R}(q) \\
     -Z^{-1}_k d^{\phi\pi}_{k,R}(q) & Z^{-1}_k d^{\phi\phi}_{k,R}(q)
   \end{matrix}
   \right),\vspace{0.0cm}
\ee
where we introduced the shorthands
\begingroup
\interdisplaylinepenalty=10000 
\begin{subequations}
\label{Eq:prop_comp}
\bea
  d^{\pi\phi}_{k,R}(q)&=&\frac{1}{\ri\omega K_k^{-1} + a_k \vec{q}^{\,2}+ \vec{q}^{\,2}\vec{q}^{\,2}_{R}}\,,\\
d^{\phi\pi}_{k,R}(q)&=&\frac{1}{-\ri\omega K_k^{-1} + a_k \vec{q}^{\,2}+  \vec{q}^{\,2} \vec{q}^{\,2}_{R}}\,,\\
  d^{\phi\phi}_{k,R}(q)&=&\frac{2 K_k^{-1} \vec{q}^{\,2}}{\omega^2 K_k^{-2} + (a_k \vec{q}^{\,2}+ \vec{q}^{\,2} \vec{q}^{\,2}_{R})^2}\,,
\eea
\end{subequations}
\endgroup
with $\vec{q}^{\,2}_{R} =  \vec{q}^{\,2} + R^{\rm L}_{k}(q)$ and $q=(\omega,\vec{q})$.
\vspace{0.2cm}
\section{Flowing couplings}
\label{Sec:flow}

\subsection{\texorpdfstring{$\beta$}{beta} functions} 

Using the Feynman rules derived in the previous section, we can express the flows of the couplings and wave function renormalizations appearing in Eq.~(\ref{Eq:ansatz}). 
Note that for all terms containing one single $\pi$ field, the $q_1$ momentum always refers to $\pi$, while $q_i$ $(i \geq 2)$ correspond to $\phi$. Using the one-loop structure of the flow equation, the following combinations are expressed in terms of diagrams:
\begin{widetext}
\begin{subequations}
\label{Eq:flows}
\bea
\label{Eq:Zpi_flow}
k\partial_k (-Z_{k} K^{-1}_k \vec{q}_1^{\,2}) &=& 0\,, \\
\label{Eq:Zphi_a_flow}
k\partial_k \big(-Z_{k} (\vec{q}_1^{\,4}+a_k \vec{q}_1^{\,2})\big)&=& k\tilde{\partial}_k \Bigg[\ \raisebox{-0.47cm}{\includegraphics[width=0.2\textwidth]{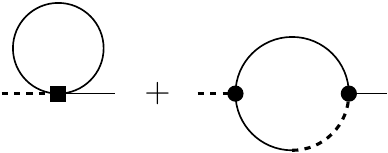}}\ \Bigg]\,,\\
\label{Eq:active_flows}
k\partial_k \Big(-\frac12 Z^\frac{3}{2}_{k} {\cal V}_k(q_1,q_2,q_3)\Big)&=& k\tilde{\partial}_k \vast[\ \raisebox{-1.0cm}{\includegraphics[width=0.5\textwidth]{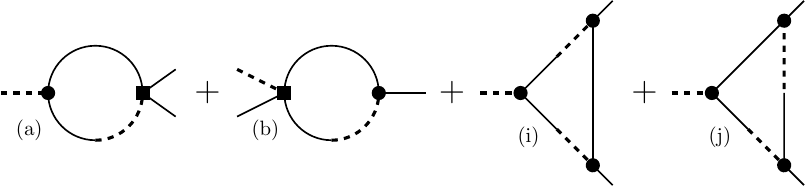}}\quad \vast]\,,\\
\label{Eq:Z4u_flow}
k\partial_k (-Z_k^2 u_k\vec{q}_1^{\,2}) &=& k\tilde{\partial}_k \vast[\ \raisebox{-1cm}{\includegraphics[width=0.5\textwidth]{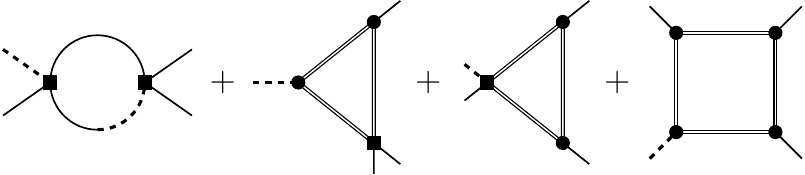}}\quad \vast]\,.
\eea
\end{subequations}
\end{widetext}

In the graphical representation of the right-hand side of the flow equations, we depart from the common FRG practice of indicating $\partial_k R_k^\textrm{L}$ as an insertion in each diagram, as in our case the $k$ derivative is understood to act on the corresponding integrand. 
As a result, there is a one-to-one formal correspondence between our diagrams and the ones occurring within the perturbative RG, which can be readily exploited in a direct comparison.

In the diagrams, black circles and squares correspond to three- and four-point vertices, respectively, derived from ${\cal P}$.
The density field $\phi$ is depicted with solid line, while the dashed one represents the response field $\pi$.
As a result, a solid line connecting two vertices corresponds to the propagator $d^{\phi\phi}_{k,R}$, while dashed-solid and solid-dashed lines to $d^{\pi\phi}_{k,R}$ and $d^{\phi\pi}_{k,R}$, respectively.
The labels (a), (b), (i), and (j) in Eq.~\eqref{Eq:active_flows} distinguish topologically different bubble and triangle diagrams to be evaluated in the Appendix.
The last three terms on the right-hand side of Eq.~\eqref{Eq:Z4u_flow} are generic diagrams in which only the fields are explicitly specified. 
For compactness, the generic propagators are shown as double lines.
When replaced with explicit propagators from Eq.~\eqref{Eq:prop_comp}, one obtains several contributions, which are discussed in Appendix.
When taking the derivative $\tilde{\partial}_k$ of the regularized propagator matrix \eqref{Eq:gamma20inv}, we will disregard its action on $Z_k$ coming from the regulator.
Last, the vertex function $\mathcal{V}_k$ is defined in Eq.~(\ref{Eq:vertex}).

In general, the contributions on the right-hand side of Eq.~\eqref{Eq:flows} carry complicated external momentum dependence, but we only need their small momentum behaviors to match the combinations present on the left-hand side.
Since the expansion in Eq.~\eqref{Eq:flow3} does not generate a contribution quadratic both in the $\pi$ field and the momentum, there is no contribution in the right-hand side of Eq.~(\ref{Eq:Zpi_flow}). 
This shows that $K_k = Z_k K$, where $K$ is the corresponding UV value of the coupling. 
Note that $K_k$ drops out from all fluctuation integrals [see Eqs.~\eqref{Eq:b3b_v2} and \eqref{Eq:w-int_b3b} for an example] and completely decouples from the RG flows of the couplings, therefore, it does not play any role in the fixed point structure to be determined.

We note that, strictly speaking, the low-momentum structure of the right-hand side of Eq.~(\ref{Eq:active_flows}) does not follow that of ${\cal V}_k$. 
Upon evaluating the diagrams, one finds that terms proportional to $\sim \phi^2\Delta \pi$ also emerge, indicating that a cubic nonlinearity is generated in the free energy at intermediate scales.
This comes as no surprise, as the active terms violate time-reversal symmetry, which in turn generates interactions in the free energy that violate the $\mathbb{Z}_2$ symmetry under $\phi \rightarrow -\phi$.
The correct interpretation is that one must start with a free energy, which explicitly violates this $\mathbb{Z}_2$ symmetry at the UV scale, so that these UV terms are precisely absorbed by the aforementioned flows as $k\rightarrow 0$. 
Note that this is not an ad hoc cancellation of inconvenient terms, but rather a necessary step to preserve the very definition of AMB+ in the IR.
In AMB+, by definition, the long-distance physics must be governed by a free energy symmetric under $\phi \rightarrow -\phi$. 
This implies that, due to the very nature of the RG, one must, at least from a mathematical point of view, compensate at the UV scale for fluctuation effects arising at intermediate scales.
Due to this construction, toward the end of the flows (as $k\rightarrow 0$) all interactions that violate the $\mathbb{Z}_2$ symmetry can be safely discarded, as their corresponding couplings vanish.
Consequently, all the $\beta$ functions to be presented are valid in the limit $k \rightarrow 0$.
Since our interest lies in finding scaling behavior, this does not affect the fixed points under investigation.

The $\beta$ functions are defined as the logarithmic $k$ derivatives (i.e., $k\partial_k$) of the dimensionless couplings
\footnote{Choosing $[K]=0$, the first three terms of the Lagrangian \eqref{Eq:Lag_rescaled} give $[\phi]=(d-2)/2$, $[\pi]=(d+2)/2$ and $[t]=-4$, which leads to $[a]=2$, $[u]=4-d$, and $[\lambda]=[\nu]=[\zeta]=(2-d)/2$.}:
\begin{align}  
\label{Eq:dimless_cpls}
\begin{split}
\bar{a}_k = a_k \,k^{-2}\,, \quad \bar{u}_k = u_k \Omega_d\,k^{d-4}\,,\\
\big(\bar{\lambda}_k,\bar{\nu}_k,\bar{\zeta}_k\big)=\big(\lambda_k,\nu_k,\zeta_k\big)\sqrt{\Omega_d}\,k^{(d-2)/2}\,.
\end{split}
\end{align}
They are introduced such that $\Omega_d=S_{d-1}/(2\pi)^d$ is eliminated from the contribution of each diagram, where $S_{d-1}\!=\!2\pi^{d/2}/\Gamma(d/2)$ is the surface area of the unit $(d-1)$-sphere.

Introducing also the anomalous dimension 
\bea
\eta_k = -\frac{k\partial_k Z_{k}}{Z_{k}}\,,
\eea
one obtains
\begin{subequations}
\label{Eq:betafunc}
\begin{align}
\beta_{a}&=(-2+\eta_k )\bar{a} + t_{a} + b_{a}\,,\\
\beta_{\nu}&=\Big(\frac{d-2}{2}+\frac32 \eta_k\Big) \bar{\nu} + b_{\nu} + \mathcal{T}_{\nu}\,,\\
\beta_{\lambda}&= \Big(\frac{d-2}{2}+\frac32 \eta_k\Big) \bar{\lambda} + b_{\lambda} + \mathcal{T}_{\lambda}\,,\\
\beta_{\zeta}&=\Big(\frac{d-2}{2}+\frac32 \eta_k\Big) \bar{\zeta} + b_{\zeta} + \mathcal{T}_{\zeta}\,,\\
\beta_{u}&=(d-4+2 \eta_k )\bar{u} + b_{u} + \mathcal{T}_{u} + \mathcal{B}_{u}\,,
\end{align}
\end{subequations}
where $t_i$, $b_i$, $\mathcal{T}_i$ and $\mathcal{B}_i$ ($i=a,\nu,\lambda,\zeta,u$) refer to the contributions of the tadpole, bubble, triangle and box diagrams, respectively [see also Eqs.~(\ref{Eq:flows})], to the coupling ``$i$.''
Introducing the shorthand notation $\Delta=(1+\bar{a})^{-1}$ for the dimensionless propagator, their respective expressions are as follows.

\vspace{0.2cm}
\noindent{\it Tadpole:}
\bea
t_a = -\frac{6 \bar{u}}{d} \Delta ^2\,.
\eea
\noindent{\it Bubble:}
\begin{subequations}
\begin{align}
b_a =& -\frac{ [(d-2) \bar{\zeta} +2 d \bar{\lambda}] \bar{\nu} }{d (4+d)} \Delta^3\,, \\
\label{Eq:bnu}
b_{\nu} =& \frac{3 \bar{u} [(4-7 d) \bar{\zeta} -2 (d-4) \bar{\lambda} +(4+7 d) \bar{\nu}] }{d^2}\Delta^3,\nonumber\\
 &+\frac{6 \bar{u} [(d-4)\bar{\zeta} +(2+d) (2 \bar{\lambda} -\bar{\nu})]}{ d (2+d)}\Delta^4\,, \\
\label{Eq:blambda}
b_{\lambda} =& \frac{3\bar{u} [(7 d-4) \bar{\zeta} +2 (d-4) \bar{\lambda} -(4+3 d) \bar{\nu}]}{d^2} \Delta^3\nonumber\\
&+\frac{6 \bar{u}\bar{\nu}}{d}\Delta^4\,, \\
b_{\zeta}=& 0\,, \\
b_u =& \frac{36 \bar{u}^2}{d} \Delta^3\,.
\end{align}
\end{subequations}

\begin{widetext}
\noindent
{\it Triangle:}
\begin{subequations}
\bea
\mathcal{T}_{\nu}&=&{\frac{1}{2 d (2+d) (4+d)}}\Big[\bar{\nu}  \big(-3 (d-2) (8+7 d)\bar{\zeta} ^2-12 [6+d (5+4 d)] \bar{\zeta} \bar{\lambda} +(d-4) (46+13 d) \bar{\zeta} \bar{\nu} \nonumber\\
&&\hspace{2.7cm}+2 (2+d) [-6 (d-6) \bar{\lambda}^2+(46+13d) \bar{\lambda} \bar{\nu} -6 d \bar{\nu} ^2] \big)\Big]\Delta^4\nonumber\\
&&-\frac{1}{d (2+d)}4 \big[(d-4) \bar{\zeta} +2 (2+d) \bar{\lambda} \big] \bar{\nu} ^2  \Delta^5,\\
\mathcal{T}_{\lambda}&=&\frac{1}{d (2+d) (4+d)} \bar{\nu}\Big(9 (d-2) (d+1) \bar{\zeta} ^2+6 [d (3+4 d)-6] \bar{\zeta} \bar{\lambda}-3 (d-2) (4+d) \bar{\zeta} \bar{\nu}\nonumber\\
&&\hspace{2.7cm}+2 (2+d) \bar{\lambda}  \big[6 (d-2) \bar{\lambda}-(13+4 d) \bar{\nu}\big]\Big)\Delta^4\nonumber\\
&&+\frac{1}{d (2+d)}2 \big[(d-2) \bar{\zeta} +2 (2+d) \bar{\lambda} \big] \bar{\nu} ^2\Delta^5\,, \\
\mathcal{T}_{\zeta}&=&-\frac{2 \bar{\zeta}\bar{\nu}\big(-3 (d-4) \bar{\zeta} +6 (4+d) \bar{\lambda} +[26+d (9+d)] \bar{\nu} \big)}{d (2+d) (4+d)}\Delta ^4+\frac{8 \bar{\zeta}\bar{\nu} ^2}{d (2+d)}\Delta^5\,,\\
\mathcal{T}_u &=& \frac{18 \bar{u} \big[(d-2) \bar{\zeta} +d (2 \bar{\lambda} -\bar{\nu} )\big] \bar{\nu}}{d (4+d)}\Delta^4\,.
\eea    
\end{subequations}
\end{widetext}
{\it Box:}
\be
\mathcal{B}_{u}=-\frac{4  [(d-2) \bar{\zeta} +2 d \bar{\lambda}] \bar{\nu} ^3 }{d (8+d)}\Delta ^5\,.
\ee    
Here, for simplicity, we dropped the $k$ subscript from all flowing couplings.

The anomalous dimension turns out to be
\bea
\label{Eq:eta}
\eta&=&\frac{1}{2d (2+d)^2} \Big((2-d) (8+7 d) \bar{\zeta} ^2
-4 [6+d (5+4 d)] \bar{\zeta} \bar{\lambda} \nonumber\\
&&\hspace{2.2cm}+(d-4) (10+3 d) \bar{\zeta}\bar{\nu} \nonumber\\
&&-2 (2+d) \left[2 (d-6)\bar{\lambda} ^2-(10+3 d)\bar{\lambda}\bar{\nu} +2 d \bar{\nu} ^2\right]\Big)\Delta ^3\nonumber\\
&&-\frac{(d-4)\bar{\zeta}+2 (2+d) \bar{\lambda}}{d (2+d)}\bar{\nu}\Delta^4\,,
\eea
which does not represent an independent flow in the sense that it can be expressed with the couplings of the theory. 
Note that $\eta$ is independent on $\bar{u}_k$, as it is derived solely from the bubble diagram in Eq.~\eqref{Eq:Zphi_a_flow}.

\subsection{Models closing under the RG \label{ss:models}}

We saw in the previous subsection that the AMB+ has five independent running couplings, of which three ($\bar{\lambda}, \bar{\nu}, \bar{\zeta}$) describe activity in the system, while the remaining two ($\bar{a}, \bar{u}$) are associated with the passive model. 
Based on the argument presented there concerning the omission of all terms that violate the $\phi \rightarrow -\phi$ symmetry of the free energy, one observes that the coupling space contains invariant subspaces under the RG flows.
This means that under the assumption of the ansatz given by (\ref{Eq:ansatz}), if one starts the flow with model parameters that belong to such a subspace, then the RG flows remain confined to it.
Having already determined the $\beta$ functions, in what follows, we review those subspaces, or in other words, classes of models that close with respect to the RG. 

{\it AMB}. The AMB+ without the $\zeta$ term leads to AMB. This model closes with respect to the RG, meaning that $\bar{\zeta}=0$ implies $\beta_{\zeta}=0$.

{\it Equilibrium models}. As discussed in Ref.~\cite{caballero2018}, setting $\bar{\zeta}=0$ and $\bar{\nu}=-2\bar{\lambda}$ leads to equilibrium models that satisfy detailed balance. 
In such situations, the chemical potential can always be obtained as the functional derivative of the free energy.%
\footnote{\label{fn:on_K1}When $\lambda=-\nu/2$, one can use $\Delta\phi^2=2\big(\vec\nabla\phi\big)^2 + 2\phi\Delta\phi$ to write $\mu_{\rm A}=\frac{\delta}{\delta \phi}\Delta\mathcal{F}[\phi]$, with $ \Delta\mathcal{F}[\phi]=-\lambda \int\rd^d x\, \phi\big(\vec\nabla\phi\big)^2$. 
Therefore, the full chemical potential can be obtained from the free energy \eqref{Eq:free_energy} by doing the shift  $K\to K(\phi)=K+2K_1\phi$, where $K_1=\nu/2.$}
This property always remains true throughout the RG flow as under the above conditions $\beta_\nu=-2\beta_\lambda$.

{\it Model B}. The AMB+ is known to inherit the fixed points of model B.
Model B is a special case of equilibrium models, where all activities vanish, $\bar{\lambda}=\bar{\nu}=\bar{\zeta}= 0$. 
As expected, for vanishing activity the $\beta$ functions of the active coupling disappear and the remaining ones correspond exactly to those of the $\phi^4$ model. 
Therefore, the static properties of a system described by model B are governed by the Wilson-Fisher fixed point.
Note that the coupling space becomes two-dimensional, spanned by $\bar{a}$ and $\bar{u}$, and that in this case $\eta=0$.

{\it cKPZ+ model}. By setting $\bar{a}=\bar{u}=\bar{\nu}=0$, we see that their corresponding $\beta$ functions also vanish, while the remaining ones, i.e., ($\bar{\lambda},\bar{\zeta})$, are those of the cKPZ+ model \cite{CaballeroPRL121} (that is the conserved Kardar-Parisi-Zhang equation supplemented with the $\zeta$ term)
\footnote{In Ref.~\cite{CaballeroPRL121}, $\lambda_k$, $\zeta_k$ and $K_k$ are the running couplings, but note that here $K_k=Z_k K$, i.e., in our description $\eta_k= -k\partial_k Z_{k}/Z_{k}$ replaces the flow equation for $K_k$.
Even so the parameter space is two-dimensional.}. The RG flow respects the $\bar{\lambda}/\bar{\zeta}$ ratio, as $\beta_\zeta/\bar{\zeta} = \beta_{\lambda}/\bar{\lambda}$. 
This relation reduces the independent flows to a single one.

{\it Case of reduced activity and zero self-interaction:} The $\beta$ functions reveal a new class of RG invariant models, resembling the cKPZ+ model. 
To our best knowledge, this class has not been reported elsewhere. 
By setting the self interaction strength $\bar{u}=0$ and imposing $\bar{\lambda}=\bar{\zeta}(2-d)/(2d)$, we get $\beta_u=0$ and $\beta_\lambda=\beta_\zeta(2-d)/(2d)$. 
That is, not only does the $\bar{u}=0$ prescription hold throughout the RG flow, but the value of the ratio $\bar{\lambda}/\bar{\zeta}$ is also preserved. 
This renders the coupling space three-dimensional, spanned by $\bar{a}$, $\bar{\nu}$, and $\bar{\lambda}$. 
As a special case, one may set $\bar{\lambda}=\bar{\zeta}=0$, leading again to an RG invariant model with a two-dimensional coupling space ($\bar{a}$, $\bar{\nu}$). 
Here the anomalous dimension turns out to be simply $\eta=-2\bar{\nu}^2/[(1+\bar{a})^3(d+2)]$. 
Note that in both cases, that is when $\bar\lambda\propto \bar\zeta\ne 0$ and when both $\bar\lambda$ and $\bar\zeta$ are set to zero, if $\bar{a}=0$ is also imposed, then $\beta_a=0$, and as a result the RG flows close.

\subsection{Universalities in the flows}

Here we discuss specific universalities, i.e., regulator independence of the $\beta$ functions obtained in the previous subsection. 
We note that all the forthcoming analyses in this subsection are only valid for $\bar{a}=0$. 
That is, it is in particular interesting from the point of view of the $\epsilon$ expansion, where $\bar{a}$ can always be set to zero, as it always only produces subleading contributions in $\epsilon$.

First, let us recall that for any $n \in\mathbb{Z}$, the following integral is regulator independent \cite{Berges:2000ew}:
\be
\label{Eq:Idn}
I(d,n) = \int \frac{\rd^dq}{(2\pi)^d} k\partial_k \frac{-1}{(\vec{q}_R^{\, 2})^n}\,,
\ee
if the dimension is set to $d=2n$. In Eq.~(\ref{Eq:Idn}), once again, $\vec{q}_R^{\, 2}=\vec{q}^{\, 2}+R_k(\vec{q})$. 
In order to prove the aforementioned statement, one makes use of the fact that any regulator, based on dimensional grounds, can be written in the form of $R_k(\vec{q})=k^2 r(|\vec{q}|/k)$, where $r(x)$ is a dimensionless profile function, being regular at $x=0$ [e.g., $r(0)=1$], while tends to zero as $x\rightarrow \infty$. 
Using the aforementioned form of $R_k(\vec{q})$, one can easily exchange the $k$ derivative to the $|\vec{q}|$ derivative:
\be
k\partial_k R_k(\vec{q}) = -|\vec{q}|\partial_{|\vec{q}|} R_k(\vec{q}) + 2R_k(\vec{q})\,,
\ee
only to realize that if $d=2n$, then the integrand of the radial integral of Eq.~(\ref{Eq:Idn}) becomes a total derivative with respect to $|\vec{q}|$, which can be trivially integrated over, without specifying the actual regulator. (Only its boundary conditions are needed.) 
After a short calculation, we get that the radial integral is exactly $1$, and arrive at
\be
I(d,d/2) = \Omega_d\,,
\ee
where, once again, $\Omega_d=\int d\Omega_d/(2\pi)^d$.
This identity has been known for a long time, but now we claim that the family of integrals
\be
I^{(\alpha)}(d,n) = \int \frac{\rd^dq}{(2\pi)^d}k\partial_k \frac{-(\vec{q}^{\, 2})^{\alpha}}{(\vec{q}_R^{\, 2})^{n+\alpha}}\,,
\ee
is also independent of the regulator and, furthermore, is independent of the value of $\alpha\in\mathbb{R}$, if $d=2n$. 

The proof is simple. Notice that the radial integral in $I^{(\alpha)}(d,n)$ is the same as in $I(d,n)$ with the substitutions $d\rightarrow d+2\alpha$ and $n\rightarrow n+\alpha$. 
That is, the integral is independent of the regulator, if $d+2\alpha=2(n+\alpha)$, but that is the same condition we had earlier, i.e., $d=2n$. 
Since the radial integral once again gives exactly $1$, we arrive at $I^{(\alpha)}(d,d/2) = I(d,d/2)=\Omega_d$.

This has important consequences for the expressions of each $\beta$ function. 
Notice that if $\bar{a}=0$, any diagram listed in the previous subsection contributes to the $\beta$ functions through specific $I^{(\alpha)}(d,n)$ integrals, with $n$ and $\alpha$ being chosen appropriately. 
For example, the contribution of the tadpole to $\beta_a$, i.e., $t_a$, appears through $\sim \int_q k\partial_k (1/\vec{q}_R^{\, 2})$, which means that it is universal for $d=2$. 
Furthermore, the contribution of the bubble, i.e., $b_a$, coming from $\sim \int_q k\partial_k (\vec{q}^{\, 4}/\vec{q}_R^{\, 4}$), is universal for $d=0$.
The same bubble diagram contributes to the wave function renormalization through the integrals $\sim \int_q k\partial_k (\vec{q}^{\, 2}/\vec{q}_R^{\, 4}$) and $\sim \int_q k\partial_k (\vec{q}^{\, 4}/\vec{q}_R^{\, 6}$), both universal in $d=2$.
One may analyze all diagrams to arrive at the conclusion that the individual contributions to $\beta_{\nu}$, $\beta_{\lambda}$, $\beta_{\zeta}$ are universal for the {\it bubble} in $d=4$ and the {\it triangle} in $d=2$. As for $\beta_u$, we see universality for the {\it bubble} in $d=4$, for the {\it triangle} in $d=2$, and for the {\it box} in $d=0$.
The flow of the anomalous dimension, solely determined by the bubble diagram, is universal for $d=2$. 
These findings are in complete agreement with the $\beta$ functions calculated using the perturbative RG, employing the $\epsilon$ expansion \cite{caballero2018}, that is, one can reproduce those results in the appropriate dimensions from Eqs.~(\ref{Eq:betafunc}) and (\ref{Eq:eta}).

\section{Fixed point structure}
\label{Sec:fixed-points}

When searching for RG fixed points of the AMB+, one encounters some that are familiar from specific models listed in Sec.~\ref{ss:models}, which contain fewer couplings. 
Some of these fixed points can be determined analytically in arbitrary dimension.

\subsection{FRG analysis}

In case of model B, we immediately find  analytically the Gaussian and the WF fixed points, the latter being
\be
\bar{a}_*=-\frac{4-d}{16-d}\,,\quad
\bar{u}_*=\frac{48 d (4-d)}{(16-d)^3}\,.
\ee 
In the $(\bar a,\bar u)$ space, the stability matrix at the Gaussian fixed point has the eigenvalues 2 and $4-d$, and hence there are two relevant directions for $d<4$. 
At the WF fixed point, there are two relevant directions for $4\le d<16$ and one relevant direction otherwise.
For both fixed points, $\eta=0$.

For the cKPZ+ model, one recalls that $\beta_\zeta/\bar{\zeta}= \beta_{\lambda}/\bar{\lambda} =(d-2+3\eta)/2$, meaning that the cKPZ+ lines of fixed points \cite{CaballeroPRL121} can also be obtained analytically in arbitrary dimensions by solving the equation $d-2+3\eta=0$ for ($\bar{\zeta}_*,\bar{\lambda}_*$) pairs.

Turning to the case of reduced activity and zero self-interaction, 
the fixed point in the $\bar a_*=\bar u_*=0$ case can be obtained analytically.
However, due to the length of the $d$-dependent expressions, we only provide its numerical values for $d=3$ in Table~\ref{tab:FP}.
When $d<2$, this fixed point is real only in a narrow range of $d$ close to $d=2$.
On the other hand, the special case, in which besides $\bar u=0$ one also has $\bar{\lambda}=\bar{\zeta}=0$, can be easily treated.
The fixed points are
\begin{subequations}
\begin{gather}
\bar a_*=-\frac{d^2+20(d+2)}{(4+d)^2}\,,\quad
\bar\nu_*=24\sqrt{3}\frac{(2+d)^2}{(4+d)^3}\,,\\
\bar a_*=0\,,\quad
\bar\nu_*=\sqrt{\frac{(4+d)(d^2-4)}{6(8+3 d)}}\,.
\end{gather}
\end{subequations}
The first fixed point has $\eta=2$ and only one relevant direction. 
For the second one, one obtains the expression $\eta=(2-d)(4+d)/[3(8+3 d)]$, and the eigenvalues of the stability matrix 
are $d-2$ and $[d^2+20(d+2)]/[3(8+3 d)]$, showing that for $d>2$ there are two relevant directions.

\begin{table}[tb!]
\centering
\caption{Fixed point values for $d=3$ in the AMB+ and related models with less couplings (N/A, i.e., not applicable, stands for the missing coupling compared to the AMB+). 
The number of relevant directions in a given model is given by $R$.
Note that the reduced coupling $\bar{a}$ always takes part in the stability analysis, even when its respective fixed point value is zero.
\label{tab:FP}}
\begin{tabular}{ >{\Centering\arraybackslash}m{1.8cm}  ccccccc }
\toprule
Model & $\bar a_*$ & $\bar u_*$ & $\bar\lambda_*$ & $\bar\nu_*$ & $\bar\zeta_*$ & $\eta_*$ & R \\
\midrule 
\multirow{5}{*}{AMB+} & $-0.68$ & 0.084 & 0.17 & 0.37 & $-0.20$ & $-1.75$ & 2 \\
 & $-0.56$ & 0.019 & 0.49 & 0.70 & $-3.2$ & 2.0 & 4 \\
 & $-0.25$ & 0.26 & 0.41 & $-0.57$ & 0.035 & $-0.41$ & 3 \\
 & 0.44 & 0.40 & 4.2 & $-5.9$ & $-23$ & 2.11 & 3 \\
\midrule 
\multirow{3}{*}{AMB} & $-0.52$ & 0.16 & 0.10 & 0.57 & N/A & $-1.24$ & 2 \\
 & $-0.21$ & 0.23 & 0.44 & $-0.51$ & N/A & $-0.25$ & 2 \\
 & 0.66 & 12.94 & 14.27 & $-4.26$ & N/A & 4.72 & 1 \\
\midrule 
\multirow{3}{*}[0pt]{\parbox{1.8cm}{Equilibrium}} & $-0.64$ & 0.14 & N/A & $-0.37$ & N/A & 0.013 & 1\\
 & $-0.35$ & 0.29 & N/A & $-0.66$ & N/A & $-0.65$ & 2\\
 & 0.024 & 0.0088 & N/A & $-0.51$ & N/A & $-0.14$ & 3\\
\midrule 
\multirow{5}{*}[0pt]{\parbox{1.8cm}{\Centering No $u$-term\\ and\\ reduced \\ activity}} & $-2.47$ & N/A & $-0.24$ & 3.29 & N/A & 2 & 1 \\
 & $-0.52$ & N/A & 0.44 & 0.73 & N/A & 2 & 2 \\
 & 0.8 & N/A & 7.02 & $-8.5$ & N/A & 2 & 2 \\
 & 0 & N/A & 0.045 & 0.66 & N/A & $-0.14$ & 3 \\
 & 0 & N/A & 0.50 & $-0.90$ & N/A & $-0.12$ & 3 \\
\bottomrule
\end{tabular}
\end{table}

The remaining fixed points of the AMB+ can be obtained only numerically, and they are listed in Table~\ref{tab:FP} for $d=3$. 
Since one also finds fixed points with special values of the couplings, in accordance with the model classification described in Sec.~\ref{ss:models}, we indicated the reduced model in which they appear. 
A reduced model contains fewer couplings,
and we show the missing ones, which would be zero or proportional to the value of another coupling in the AMB+, as N/A (not applicable). 
The number of relevant directions is always calculated within the indicated model.
In some cases, this number would be different if the stability of the fixed point was analyzed in the full AMB+.
It is worth mentioning that the reduction of the AMB+ is also shown by the behavior of the eigenvectors of the $5\times5$ stability matrix.
For example, in the case of the equilibrium fixed points, defined by the relation $\bar\nu_*=-2\bar\lambda_*$, it turns out that the projections of three eigenvectors onto the $(\bar\lambda, \bar\nu)$ plane lie precisely along the line satisfying the equation $\bar\nu=-2\bar\lambda$.

\subsection{Remarks on the perturbative RG \label{ss:mom-shell_RG}} 

The diagrammatic approach, introduced in the previous section, can also be used to derive the $\beta$ functions corresponding to the perturbative RG, which was also used in Ref.~\cite{caballero2018} to obtain results in the $\epsilon$ expansion. 
In order to derive the $\beta$ functions in this framework, one needs to evaluate all diagrams in Eq.~(\ref{Eq:flows}) in a momentum shell $|\vec{q}|\in [\Lambda/b,\Lambda]$, while formally erasing everywhere the $k\tilde{\partial}_k$ operator. 
Then, by setting $b=1+\rd b$, and taking the negative derivative with respect to $\rd b$, after rescaling all couplings with $b$, in accordance with their dimensionality (including the anomalous dimension), one gets the corresponding $\beta$ functions.%
\footnote{Note that, in standard textbook analyses, the $\beta$ functions are defined with an additional minus sign compared to our work.}
The structure is identical to that of Eq.~(\ref{Eq:betafunc}), but the corresponding expressions for the tadpole, bubble, triangle and box diagrams are different. 
Note, however, that the universality explained in the previous section does hold, which is due to the fact that the standard perturbative RG results can also be reproduced within the FRG framework by using the Wilsonian regulator, $R_k^{\rm W}(q)=\lim_{M\rightarrow \infty} M^2 \Theta(k^2-\vec{q}^{\, 2})$. 
Note that the perturbative approach requires that both $(4-d)$ and $(d-2)$ must be treated as an expansion parameter, raising doubts on whether in $d=3$ the results are trustworthy. 
In the FRG scenario, being nonperturbative, no such requirements need to be imposed.

In the case of the perturbative RG, one encounters the same classes of fixed points and RG invariant models as in the FRG, with the exception of the class of fixed points in which $\bar{u}_*=0$ and $\bar{a}_*\ne0$. 
The reason for this is as follows. 
For $\bar{u}_*=0$, the condition $\beta_u=0$ requires $\bar{\lambda}=\bar{\zeta}(2-d)/(2d)$, just as in the FRG, and since $\eta$\footnote{This corresponds to $M$ in the notation of Ref.~\cite{caballero2018}.} is independent of $\bar{a}$, the $\beta$ function for $\bar{a}$ is of the form $\beta_a=\bar{a} f(\bar{\lambda},\bar{\nu})$.
Therefore, $\beta_a=0$ admits a nontrivial solution with $\bar{a}\ne 0$ only if the function $f$ vanishes at the solution of the system of equations $\{\beta_\lambda=0$, $\beta_\nu=0\}$, which is also independent of $\bar{a}$. 
However, it turns out that $f\ne0$ at the solution of the above system, and as a result, this class of fixed points, obtained in the FRG case, i.e., having $\bar{u}=0$ and $\bar{a}\ne0$, is not present in case of the perturbative RG.

\subsection{Phase portrait of the equilibrium model}

\begin{figure}[!tb]
\centering
\includegraphics[width=0.45\textwidth]{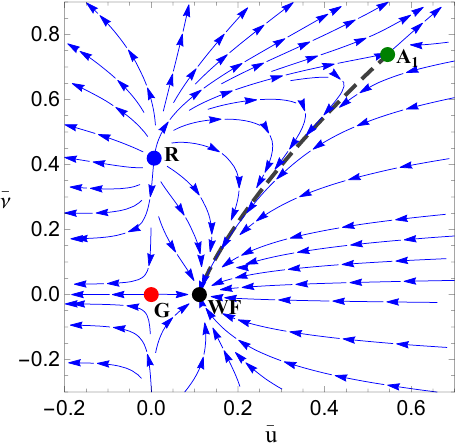}
\caption{Two-dimensional phase portrait of the equilibrium model obtained using the perturbative RG.
Each fixed point has a relevant direction perpendicular to the plane, that is along the $\bar a$ axis.
The dashed line in gray indicates a separatrix. \label{Fig:MBK1-RG-2d}}
\end{figure}

In case of the perturbative RG, the fixed point structure of the equilibrium model can be visualized in a two-dimensional phase portrait because the flow of $\bar a$ decouples from the flows of the other two couplings.
This is shown in Fig.~\ref{Fig:MBK1-RG-2d}.  
We see that for small values of $\nu$, the IR behavior of the system is determined by the WF or the Gaussian fixed points, for which $\bar\nu_*=0$.
When $\bar\nu$ is larger than the value it takes at the completely repulsive fixed point R, shown as a blue blob, one can tune the couplings such that the IR behavior is determined by the fixed point A$_1$ shown as a green blob.
This could indicate the possibility of a new phase transition, not governed by the WF fixed point.
We recall that, as explained in footnote~\ref{fn:on_K1}, the equilibrium model corresponds to model B with a $\phi$-dependent $K$ of the form $K(\phi)=K+2 K_1\phi$, where $2K_1=\nu$.

\begin{figure}[!tb]
\centering
\includegraphics[width=0.45\textwidth]{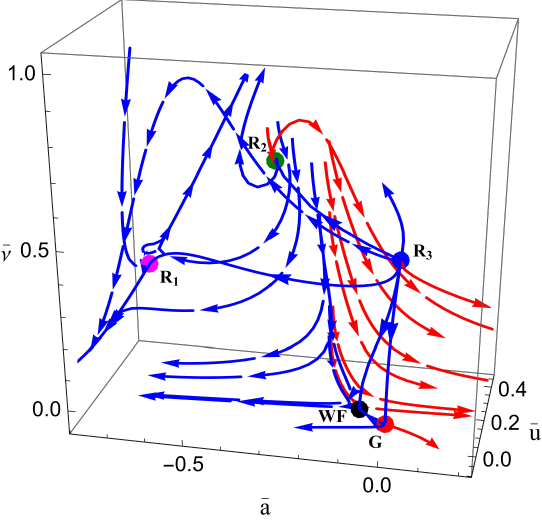}
\caption{Three-dimensional phase portrait of the equilibrium model within the FRG.
Compared to the perturbative RG, a new fixed point appears, namely R$_1$, shown as a magenta blob. The indices $i$ of the fixed points R$_i$ refers to the number of relevant directions indicated in Table~\ref{tab:FP}. The flows shown in red go to regions of the parameter space where $\bar a>0$. \label{Fig:MBK1-FRG}}
\end{figure}

When using the FRG approach, the flow of $\bar a$ is coupled to those of
$\bar u$ and $\bar \nu$.  
In this case, one obtains an additional fixed point compared to the perturbative RG.  
This new fixed point has the smallest value of $\bar a_*$ and it is indicated by a magenta blob in the three-dimensional phase portrait shown in Fig.~\ref{Fig:MBK1-FRG}.  
This has one attractive direction, and therefore, it prevents some flow trajectories from running to infinitely large values of the couplings, as happened when the perturbative RG was used.
In the FRG case, we also see the possibility of new phases for large values of $\bar \nu$, similarly to the perturbative case. 

\subsection{Findings in AMB+}

In the full theory of AMB+, given the large number of independent coupling constants, we do not attempt to visualize the flows. 
Any two-dimensional or three-dimensional projections of the flow chart inevitably contain flows that cross each other, making an intuitive understanding difficult.

First, we recall that in the perturbative RG study, performed in Ref.~\cite{caballero2018}, only those fixed points that merged into the WF fixed point as $d\to 2^+$ were considered to be reliably captured by the method used.
Two such fixed points were found.  
One of them, denoted $F_{\rm eq}$, is an equilibrium fixed point characterized by $\bar\zeta_*=0$ and $\bar\nu_*=-2\bar\lambda_*$. 
The other, denoted $F_4$, is a genuine AMB+ fixed point, conjectured to control the region of the parameter space, where the transition from bulk separation to microphase separation occurs in active systems.  
It was found that $F_4$ is bicritical, having two relevant directions, one in the direction of $\bar a$ and another one spanned by a linear combination of all couplings.

Using the FRG method, we also find a fixed point with two relevant directions; see the first line in Table~\ref{tab:FP}. 
This corresponds to $F_4$ of Ref.~\cite{caballero2018}, although the actual directions of relevance are different. 
This is no surprise, as only in the perturbative case does one find one of the relevant directions corresponding always to $\bar{a}$. 
With the FRG, one finds that the active couplings of $F_4$ do not vanish simultaneously for $d\to2^+$, in contrast to the situation reported in the perturbative case. 
This is not a coincidence, as the $d$ dependence of all fixed points obtained in the FRG genuinely shows a rather different scenario compared to that of those found via the perturbative RG.  
Most importantly, it turns out that none of the fixed points in the FRG approaches the WF fixed point as $d\to 2^+$.  
However, interestingly, the value of $\bar u_*$ of the first two AMB+ fixed points and the first AMB fixed point in Table~\ref{tab:FP} does get close to the value taken at the WF fixed point as $d\to 2^+$.

The FRG method predicts a different global flow chart around the $F_4$ fixed point compared to that of the perturbative study. 
After identifying the eigenvectors of $F_4$, we investigated the flows in the two-dimensional subspace spanned by the relevant directions in order to decide whether there exists a path toward the WF fixed point. 
What we see is that this is not the case. 
We systematically initialized flows on the aforementioned two-dimensional plane while staying close to $F_4$, but we found no signs of hitting any fixed point, including the WF fixed point.
What we see is that one of the eigendirections divides the plane into two disjoint regions: The flows either go to strong coupling or hit a singularity characterized by $\bar{a}=-1$ and vanishing higher couplings. 
The reason this is a singular point is that $\bar{a}=-1$ corresponds to the zero of the dimensionless inverse propagator, meaning that all flows break down at this point. 
Note that this is a genuinely FRG phenomenon, as in case of perturbation theory, $\bar{a}$ must always be dropped in all propagators based on consistency of the expansion. 

Given that no signature of an IR stable fixed point is found at strong coupling, our results show no sign that the microphase separation transition would be of second order. 
The absence of a critical point hints that it is presumably of first order. 
As for the bulk phase separation transition, present at low activities, according to Ref.~\cite{caballero2018}, is still governed by the WF fixed point. 
In the present study, however, since $F_4$ does not appear to be connected to the WF fixed point, we cannot rule out that the nature of the bulk phase transition changes for intermediate or even small activities. 
One understands this difference between the perturbative and functional approaches by realizing that in the FRG, the flow of the $\bar{a}$ parameter (corresponding to the temperature in model B) always remains in the flows of all higher-order couplings and, therefore, it cannot be considered as a quantity that is separated from the other flows. 
Since $\bar{a}$ deviates fast from its corresponding $F_4$ value, it forces all the other couplings to either diverge or vanish, leaving no room to reach the WF fixed point. We believe it is an artifact of perturbation theory that the flow of $\bar{a}$ does not matter when calculating scale dependence of the higher-order couplings, and this is why the WF fixed point seems to be connected with $F_4$.

We point out that the aforementioned results might be specific to our ansatz for the FRG flows, where only those couplings that are also present at the UV scale were considered. 
It would be interesting to go beyond the present approximation to investigate the robustness of the obtained results.

\section{Conclusions \label{Sec:concl}}

In this paper, we investigated the RG flows of AMB+ using the functional variant of the RG. To this end, we formulated AMB+ as a local field theory via MSR construction. Once an action functional is given, application of the FRG becomes straightforward. We employed an ansatz for the effective action, which treats all couplings in the original model as scale-dependent quantities, and we investigated whether they can show scaling behaviors. All $\beta$ functions were calculated using Litim's regulator function.
However, we pointed out that in specific dimensions, individual contributions to the flows show universal behavior, as their scale dependence is independent of the choice of the infrared regulator.

The $\beta$ functions allowed us to investigate the existence of subclasses of AMB+, where the RG flows close with respect to specific initial conditions. 
In addition to model B itself, we distinguished four subclasses, of which one could only be revealed by the FRG treatment, as it is beyond reach in the perturbative RG. 
We calculated all fixed points with possible physical relevance, and confirmed the existence of the bicritical $F_4$ fixed point, conjectured in Ref.~\cite{caballero2018} to be responsible for the transition from bulk phase separation to microphase separation.

In agreement with perturbation theory, we have found that for large activities, the RG flows do not hit any infrared stable fixed point and diverge to strong coupling. 
This suggests that the microphase separation transition is first order, in which physical observables exhibit discontinuous jumps with respect to the appropriate control parameters.
In contrast to the perturbative RG, however, we have found that the $F_4$ fixed point is not connected to the WF fixed point.\footnote{For the sake of completeness, we mention that we have verified that the only nontrivial fixed point connected to the WF fixed point is the equilibrium one with three relevant directions.
Instead, initializing flows close to $F_4$ at lower activities leads to a singular point where all couplings vanish except for the quadratic one ($\bar{a} \rightarrow -1$). 
This indicates that, contrary to the findings of Ref.~\cite{caballero2018}, the bulk phase separation observed at intermediate activities may not belong to the same universality class as the equilibrium phase transition of passive model B. 
This could explain why, in Refs.~\cite{siebert2018,dittrich2021}, the measured static critical exponents differ from those of the liquid-gas transition. 
Moreover, one might argue that the bulk phase separation itself becomes first order at intermediate activities. 
As expected from dimensional analysis, for perturbatively small activities the WF fixed point is attractive in all active directions. 
Therefore, in such systems, the bulk phase separation indeed lies in the same universality class as the passive model. 
Note that, in both scenarios, the activity becomes irrelevant at large scales, implying that time-reversal symmetry is effectively restored.}

There are several ways to improve the present study. First and foremost, the FRG technique, being nonperturbative, allows for the investigation of the flow of the effective action without requiring a Taylor expansion in terms of the field variables or their appropriate combinations [see the individual terms appearing in the MSR Lagrangian~(\ref{Eq:Lag})]. In principle, after identifying all combinations of the density and response fields that can carry activity, one could construct an effective action that is a nonperturbative functional of these building blocks. A simplified version of this idea would also be worth exploring, in which only the equilibrium free energy is promoted to a nonperturbative function of the density field. Furthermore, while all field rescaling factors were treated as constants in this study, they could, in principle, depend on the density field. We also note the possibility of calculating the dynamical critical exponent $z$ and studying the backreaction of the anomalous dimensions on the flow by keeping the $k$ derivative of $Z_k$ present in the regulator ${\cal R}_k$, both of which were not addressed in this work. These ideas will be pursued in future studies.

\section*{Acknowledgements}
The authors thank Ryo Hanai for suggesting this problem to us.
We also thank Fernando Caballero and Cesare Nardini for useful correspondence and Kyosuke Adachi and Hiroyoshi Nakano for discussions.
G.\,F. was supported by the Hungarian National Research, Development, and Innovation Fund under Project No.~FK142594.
N.\,Y. was supported by JSPS KAKENHI Grant No. JP24K00631 and the Ishii-Ishibashi Fund (Keio University Grant for Early Career Researchers).

\appendix*

\section{Evaluation of diagrams \label{app:graphs}}

\subsection{Contribution of the diagrams}

Here we give the contributions of the diagrams appearing in Eq.~\eqref{Eq:flows}. 
Incoming momenta $q_i$ (with $i=1,2,\dots$) are assigned to the external legs, starting with the dashed line (the $\pi$ field) and proceeding counterclockwise.
For the three-point vertices, we note that the first momentum belongs to the $\pi$ field, and that an incoming (outgoing) momentum $q$ appears with a positive (negative) sign.
The loop momentum $q$ is assigned to the plain line.
If a momentum $q'$ points from the dashed line towards the solid line then one has $d_{k,R}^{\pi \phi}(q')$, while for the opposite orientation one has $d_{k,R}^{\phi \pi}(q')$.
The wave function renormalization factors from the interaction part \eqref{Eq:P} and propagator matrix \eqref{Eq:gamma20inv} generate through Eq.~\eqref{Eq:flow3} the factor $Z_k^{n/2}$ for a diagram with $n$ external legs.

The contribution of the diagrams are denoted by letters: $t$, stands for tadpole, $b$ for bubble, $\mathcal{T}$ for triangle, and $\mathcal{B}$ for box. 
The numerical part in the subscript of these letters (e.g., ``2'' in $t_2$) indicates the number of external legs of the corresponding diagram. 
When a letter also appears in the subscript (e.g., ``a'' in $b_{3a}$), it refers to the label of the corresponding diagram.

The contributions of the tadpole and bubble diagrams in Eq.~\eqref{Eq:flows} with various number of external are
\begin{eqnarray}
t_2&=&-3\vec{q}_1^{\,2}\int_q\! Z_k u_k d_{k,R}^{\phi\phi}(q),\nonumber\\
b_2&=&\int_q\! Z_k \mathcal{V}_k(q_1,q',q) d_{k,R}^{\pi\phi}(q') \mathcal{V}_k(-q',-q,-q_1) d_{k,R}^{\phi\phi}(q),\nonumber\\
b_{3a}&=&-3\int_q\! Z_k^\frac{3}{2}\mathcal{V}_k(q_1,q',q) d_{k,R}^{\pi\phi}(q') \pvecs{q} u_k d_{k,R}^{\phi\phi}(q),\nonumber\\
b_{3b}&=&-6\vec{q}_1^{\,2} \int_q\! Z_k^\frac{3}{2} u_k d_{k,R}^{\pi\phi}(q'') \mathcal{V}_k(-q'',-q,q_3) d_{k,R}^{\phi\phi}(q),\nonumber\\
b_4&=&18 \vec{q}_1^{\,2} \int_q\! Z_k^2 u_k^2 \ppvecs{q} d_{k,R}^{\pi\phi}(q'') d_{k,R}^{\phi\phi}(q),
\end{eqnarray}
where $q'=-q-q_1$ and $q''=-q-q_1-q_2$.

The contributions of the triangle diagrams in Eq.~\eqref{Eq:flows} with three external legs are given by
\begin{subequations}
\begin{alignat}{3}
&\mathcal{T}_{3i} = \frac{1}{2}\int_q\ &&\big[Z_k^\frac{3}{2}\, \mathcal{V}_k(q_1,-q',-q'') d_{k,R}^{\phi\pi}(q') \mathcal{V}_k(q',q_2,q)\nonumber \\
&&&\times d_{k,R}^{\phi\phi}(q)\mathcal{V}_k(q'',q_3,-q) d_{k,R}^{\phi\pi}(q'')\big]\,,\\
&\mathcal{T}_{3j} =\ \int_q\ &&\big[Z_k^\frac{3}{2}\, \mathcal{V}_k(q_1,q,q') d_{k,R}^{\pi\phi}(q') \mathcal{V}_k(-q',q_2,-q'')\nonumber\\
&&&\times d_{k,R}^{\phi\pi}(q'')\mathcal{V}_k(q'',q_3,-q) d_{k,R}^{\phi\phi}(q)\big]\,,
\end{alignat}  
\end{subequations}
where $q''=q-q_3$, while $q'=-q-q_2$ in the first contribution and $q'=-q-q_1$ in the second one.

\begin{figure}[!tb]
\centering
\includegraphics[width=0.3\textwidth]{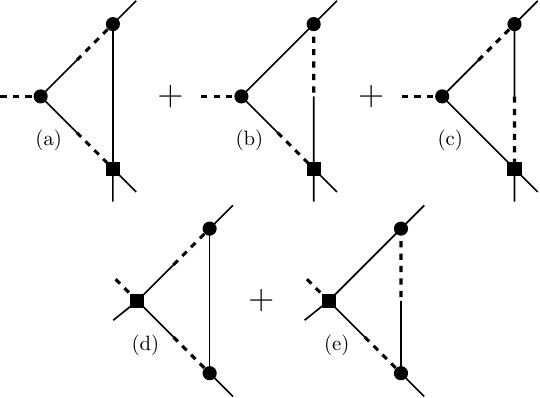}
\caption{The diagrams in the first (second) row are obtained from the first (second) generic triangle diagram of Eq.~\eqref{Eq:Z4u_flow}. 
The fields $\phi$ and $\pi$ are depicted with solid and dashed lines, respectively. 
A solid line connecting two vertices corresponds to the propagator $d^{\phi\phi}_{k,R}$, a dashed-solid line to $d^{\pi\phi}_{k,R}$ and a solid-dashed line to $d^{\phi\pi}_{k,R}$. \label{Fig:triangles}}
\end{figure}

The first generic triangle diagram with four external legs in Eq.~\eqref{Eq:Z4u_flow} results in the upper three diagrams shown in Fig.~\ref{Fig:triangles}.
Their contributions are
\begin{subequations}
\begin{alignat}{3}
  &\mathcal{T}_{4a} = -3\int_q &&\big[Z_k^2\, \mathcal{V}_k(q_1,q',q'') d_{k,R}^{\pi\phi}(q')\,u_k\pvecs{q}\nonumber \\
  &&&\times d_{k,R}^{\phi\phi}(q)\mathcal{V}_k(-q'',q_4,q) d_{k,R}^{\pi\phi}(q'')\big]\,,\\
  &\mathcal{T}_{4b} = -3\int_q &&\big[Z_k^2\, \mathcal{V}_k(q_1,q,-q') d_{k,R}^{\phi\pi}(q')\,u_k\pvecs{q}\nonumber \\
  &&&\times d_{k,R}^{\phi\pi}(q'')\mathcal{V}_k(q'',q_4,-q) d_{k,R}^{\phi\phi}(q)\big]\,,\\
  &\mathcal{T}_{4c} = -3\int_q &&\big[Z_k^2\, \mathcal{V}_k(q_1,q,-q') d_{k,R}^{\phi\phi}(q)\,u_k\ppvecs{q}\nonumber \\
  &&&\times d_{k,R}^{\phi\pi}(q'')\mathcal{V}_k(q',q_4,-q'') d_{k,R}^{\phi\pi}(q')\big]\,,
\end{alignat}
respectively. Here, $q'=-q-q_1-q_4$ in the first contribution and $q'=q+q_1$ in the last two.
Additionally, $q''=q+q_4$ in the first contribution, $q''=q-q_4$ in the second, and $q''=q+q_1+q_4$ in the third. 

The contributions of the lower two diagrams of Fig.~\ref{Fig:triangles} obtained from the second generic triangle diagram with four external legs in Eq.~\eqref{Eq:Z4u_flow} are
\begin{alignat}{3}
  &\!\!\!\mathcal{T}_{4d} = -3\vec{q}_1^{\,2}\int_q &&\big[Z_k^2\, u_k d_{k,R}^{\pi\phi}(q'')\mathcal{V}_k(-q'',q_3,-q) \nonumber \\
  &&&\times d_{k,R}^{\phi\phi}(q) \mathcal{V}_k(-q',q_4,q) d_{k,R}^{\pi\phi}(q')\big]\,,\\
  &\!\!\!\mathcal{T}_{4e} = -6\vec{q}_1^{\,2}\int_q &&\big[Z_k^2\, u_k d_{k,R}^{\phi\phi}(q)\mathcal{V}_k(-q'',q_4,-q) \nonumber \\
  &&&\times d_{k,R}^{\pi\phi}(q'') \mathcal{V}_k(-q',q_3,q'') d_{k,R}^{\pi\phi}(q')\big]\,,
\end{alignat}
where $q'=q+q_4$ and $q''=-q+q_3$ for the first contribution, while $q'=-q+q_3+q_4$ and $q''=-q+q_4$ for the second.
\end{subequations}

\begin{figure}[!tb]
\centering
\includegraphics[width=0.3\textwidth]{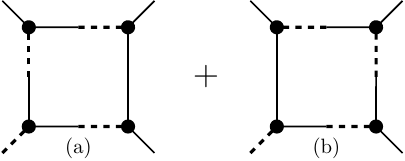}
\caption{Box diagrams obtained from the corresponding generic diagram in Eq.~\eqref{Eq:Z4u_flow}.
See Fig.~\ref{Fig:triangles} for the convention on the lines.
\label{Fig:Boxes}}
\end{figure}

Last, the contributions of the box diagrams of Fig.~\ref{Fig:Boxes} to Eq.~\eqref{Eq:Z4u_flow} are
\begin{alignat}{1}
  &\mathcal{B}_{4a}=\!\int_q\!\big[Z_k^2\,\mathcal{V}_k(q_1,-q',-q''') d_{k,R}^{\phi\pi}(q')\mathcal{V}_k(q',q_2,q) d_{k,R}^{\phi\phi}(q)\nonumber\\
  &\times \mathcal{V}_k(q'',q_3,-q) d_{k,R}^{\phi\pi}(q'')\mathcal{V}_k(q''',q_4,-q'') d_{k,R}^{\phi\pi}(q''')\big],\!\!\!\\
  &\mathcal{B}_{4b}=\!\int_q\!\big[Z_k^2\,\mathcal{V}_k(q_1,q,-q') d_{k,R}^{\phi\pi}(q')\mathcal{V}_k(q',q_2,-q'') d_{k,R}^{\phi\pi}(q'')\nonumber\\
  &\times \mathcal{V}_k(q'',q_3,-q''') d_{k,R}^{\phi\pi}(q''')\mathcal{V}_k(q''',q_4,-q) d_{k,R}^{\phi\phi}(q)\big],\!\!  
\end{alignat}
where $q'=-q-q_2$, $q''=q-q_3$, and $q'''=q-q_3-q_4$ for the first contribution, and  $q'=q+q_1$, $q''=q+q_1+q_2$, and $q'''=q-q_4$ for the second. 

\subsection{General method and the evaluation of \texorpdfstring{$b_{3b}$}{b3b}}

The method, which is applicable to all our diagrams, involves performing first the frequency integral using contour integration and then taking the $k$ derivative and shifting the loop momentum in some terms in order to produce $\partial_k R_k^{\rm L}(\vec{q})$ in all numerators, followed by expanding the integrand for small external momenta, and finally performing the angular integrals and the integral over the magnitude of the loop momentum.

The evaluation of diagrams with three external legs is the most challenging, as they depend on two external momenta. 
In contrast, with four-legged diagrams, we can choose to set two external momenta to zero, which significantly simplifies their evaluation. 
As for the two-legged bubble diagram, this can also be evaluated using the method presented in Appendix~B of  Ref.~\cite{Kobayashi:2019sus} for the calculation of the wave function renormalization constant.

To illustrate the method used for calculation, here we present in some detail the evaluation of the second three-legged bubble diagram in Eq.~\eqref{Eq:active_flows}. 
After the shift $q\to -q$ it reads
\begin{align}
\label{Eq:b3b_v2}
b_{3b}=&-12 \vec{q}_1^{\,2}\int \frac{\rd^d q}{(2\pi)^d} \bigg[Z_k^\frac{3}{2} u_k \vec{q}^{\,2} \mathcal{V}_k(-q-q_3,q,q_3)\nonumber\\
&\times \int_{-\infty}^\infty \frac{\rd\omega}{2\pi}\frac{K_k^{-1}}{(K_k^{-2}\omega^2+A^2)(\ri K_k^{-1}\omega + B)}\bigg]\,,
\end{align}
where $A=\vec{q}^{\,2}\big(a_k+\vec{q}^{\,2}_R\big)$ and $B=(\vec{q}+\vec{q}_3)^2\big[a_k+(\vec{q}+\vec{q}_3)^2_R\big]$.

Setting all external frequencies to zero, the shift $\omega\to \omega K_k$ eliminates $K_k^{-1}$ from  Eq.~\eqref{Eq:b3b_v2}, and the frequency integral gives for $A>0$ and $B>0$
\be
\label{Eq:w-int_b3b}
\int_{-\infty}^\infty \frac{\rd\omega}{2\pi}\frac{1}{(\omega^2+A^2)(\ri\omega + B)} = \frac{1}{2A(A+B)}\,.
\ee
Using the explicit form of the regulator, it is easy to check that the above two conditions are satisfied for $|\bar a_k|<1$, a condition that has to be imposed in order to avoid a singularity in the remaining momentum integral. 

As mentioned below Eq.~\eqref{Eq:flows}, the $k$ derivative has to be taken inside the integral. Then, from Eq.~\eqref{Eq:w-int_b3b} this gives
\be
\label{Eq:dk_w-int_b3b}
\tilde \partial_k \frac{1}{2A(A+B)} = f(A,B)\partial_k A +  g(A,B)\partial_k B\,,
\ee
where $f(A,B)=-(2A+B)/[2A^2(A+B)^2]$, $g(A,B)=-1/[2A(A+B)^2]$, and $\partial_k A=2 k\vec{q}^{\,2}\Theta(k^2-\vec{q}^{\,2})$.
In the second term of Eq.~\eqref{Eq:dk_w-int_b3b} we  shift $\vec{q}\to -\vec{q}-\vec{q}_3$, which results in the interchange $A\leftrightarrow B$, and therefore in the right-hand side of the flow equation \eqref{Eq:active_flows} we obtain 
\begin{align}
\label{Eq:b3b_v3}
  &k\tilde \partial_k b_{3b}=-24 k^2 u_k Z_k^\frac{3}{2} \vec{q}^{\,2}_1\!\int\frac{\rd^d q}{(2\pi)^d}\vec{q}^{\,2}\Theta(k^2-\vec{q}^{\,2}) \nonumber\\
  &\qquad \times \Big[\vec{q}^{\,2} f(A,B)\mathcal{V}_k(-q-q_3,q,q_3) \nonumber\\ 
  &\qquad\ \ \ \ + (\vec{q}+\vec{q}_3)^2 g(B,A) \mathcal{V}_k(q,-q-q_3,q_3)
\Big]\,.
\end{align}

Adopting the spherical coordinate system used in Ref.~\cite{caballero2018}, the loop momentum $\vec{q}$ is given as
\be
\vec{q}= |\vec{q}_{\,}|\begin{pmatrix}
\sin\theta\dots\sin\varphi_{d-3}\sin\varphi_{d-2} \\
\sin\theta\dots\sin\varphi_{d-3}\cos\varphi_{d-2} \\
\vdots\\
\sin\theta\cos\varphi_1\\
\cos\theta
\end{pmatrix}\,,
\ee
with $\theta,\varphi_1,\dots,\varphi_{d-3}\in[0,\pi]$, and $\varphi_{d-2}\in [0,2\pi]$.
Choosing $\vec{q}_1$ along the $x_d$ axis and $\vec{q}_3$ in the $x_d-x_{d-1}$ plane, as 
$\vec{q}_3^{\,\rm T}=|\vec{q}_3|(0, \dots, \sin\psi, \cos\psi)$ with $\psi\in[0,\pi]$, the scalar products $\vec{q}\cdot \vec{q}_1$, $\vec{q}\cdot \vec{q}_3$ and $\vec{q}_1\cdot \vec{q}_3$ can be easily expressed.
One can perform the angular integrals in Eq.~\eqref{Eq:b3b_v3}, except for those over $\theta$ and $\varphi_1$, and one obtains
\be
\int \frac{\rd\Omega_d}{(2\pi)^d}=\frac{d-2}{2\pi} \Omega_d \int_0^\pi \rd\theta \int_0^\pi \rd\varphi_1 \sin^{d-2}\theta \sin^{d-3}\varphi_1\,,
\ee
where $\Omega_d$ is given below Eq.~\eqref{Eq:dimless_cpls}.

In order to perform the integrals over the angles $\theta$ and $\varphi_1$, and over the magnitude $|\vec{q}_{\,}|$ of the loop momentum, one has to work first on the expression of $B$ given below Eq.~\eqref{Eq:b3b_v2}. 
Note that the expression of $A$ simplifies due to the presence of $\Theta(k^2-\vec{q}^{\,2})$ in  Eq.~\eqref{Eq:b3b_v3} to $A=\vec{q}^{\,2}(a_k+k^2)$, as $\vec{q}^{\,2}_R = k^2$ in the domain of integration.
Expanding for small $\vec{q_3}$, one obtains
\be
\label{Eq:reg_mom_exp}
\begin{aligned}
&(\vec{q}+\vec{q}_3)^2_R=\vec{q}_R^{\,2} + s\big[\Theta(\vec{q}^{\,2}-k^2) - (k^2-\vec{q}^{\,2})\delta(k^2-\vec{q}^{\,2})\big] \\
&\quad+\frac{s^2}{2} \big[2 \delta(k^2-\vec{q}^{\,2}) + (k^2-\vec{q}^{\,2})\delta'(k^2-\vec{q}^{\,2})\big] + \mathcal{O}(s^3)\,,
\end{aligned}
\ee
where $s=\vec{q}_3^{\, 2}+2\vec{q}\cdot\vec{q}_3$.
Since the integrand in Eq.~\eqref{Eq:b3b_v3} is also expanded in the external momenta, the coefficient of $s$ in Eq.~\eqref{Eq:reg_mom_exp} will be multiplied by $\Theta(k^2-\vec{q}^{\,2})$, and as a result the terms $\propto s$ do not contribute when integrated over the loop momentum.
As for the term $\propto s^2$, we first use the identity $(k^2-\vec{q}^{\,2})\delta'(k^2-\vec{q}^{\,2}) = -\delta(k^2-\vec{q}^{\,2})$, and then apply the Morris lemma~\cite{Morris:1994}
\[
\Theta(x) \delta(x) = \delta(x) \int_0^1 \rd t\, t = \frac{1}{2} \delta(x)\,,
\]
to obtain for some function $h(q)$, where $q=|\vec{q}_{\,}|$, 
\[
\int_0^\infty \rd q\, h(q) \Theta(k-q)\delta(k^2-q^2) = \frac{h(k)}{4 k}\,.
\]

All expansions, algebraic manipulations, and integrations discussed above can be conveniently performed using a symbolic computation program, such as \textit{Mathematica}. 
Applied to Eq.~\eqref{Eq:b3b_v3} they give
\be
\label{Eq:b3b_res}
k\tilde \partial_k b_{3b}=C_2 \vec{q}^{\,2}_1 + C_4 \vec{q}^{\,2}_1 \vec{q}^{\,2}_3\,,
\ee
where 
\begin{subequations}
\begin{align}
\!\!\!C_2=&\frac{12 k^{5+d} Z_k^\frac{3}{2}u_k\nu_k\Omega_d}{(2+d) (k^2+a_k)^3},\\
\!\!\!C_4=&-\frac{3 k^{3+d} Z_k^\frac{3}{2}u_k\nu_k\Omega_d}{d^2 (k^2+a_k)^3}\bigg(\frac{2 d\, k^2\nu_k}{k^2+a_k} \nonumber \\
\!\!\!&+\big[2(d-4)\lambda_k-(4+3 d)\nu_k - (4-7 d)\zeta_k \big]\bigg).\!\!
\end{align}
\end{subequations}

As a final step, we need to match the momentum dependence of Eq.~\eqref{Eq:b3b_res}, that is, the right-hand side of the flow equation \eqref{Eq:active_flows}, with the momentum dependence of the vertex function on its left-hand side. 
Also, since the vertex function is symmetric under $\vec{q}_2\leftrightarrow \vec{q}_3$, we have to symmetrize our result \eqref{Eq:b3b_res} and exploit momentum conservation, using $\vec{q}_2=-\vec{q}_1-\vec{q}_3$.
Then the vertex function takes the form:
\be
\label{Eq:vertex_q2_elim}
\begin{aligned}
 \mathcal{V}_k(q_1,-q_1-q_3,q_3)&=\nu_k\vec{q}_1^{\,4}-2\lambda_k\vec{q}_1^{\,2}(\vec{q}_1\cdot \vec{q}_3 + \vec{q}_3^{\,2})\\
  &\quad +\zeta_k[2(\vec{q}_1\cdot \vec{q}_3)^2 + \vec{q}_1^{\,2}(\vec{q}_1\cdot \vec{q}_3 - \vec{q}_3^{\,2})]\,,
\end{aligned}
\ee
while the symmetrized version of our result becomes
\be
\label{Eq:b3b_res_sym}
k\tilde \partial_k b_{3b}=C_2 \vec{q}^{\,2}_1 + \frac{C_4}{2} \vec{q}^{\,4}_1 + C_4
\vec{q}^{\,2}_1\big(\vec{q}_1\cdot \vec{q}_3 +\vec{q}^{\,2}_3)\,.
\ee
We must drop the term quadratic in $\vec{q}_1$, as it has no counterpart in the vertex function [see the related discussion in the paragraph preceding Eq.~\eqref{Eq:dimless_cpls}]. 
Matching the quartic parts in Eqs.~\eqref{Eq:b3b_res_sym} and \eqref{Eq:active_flows}, using Eq.~\eqref{Eq:vertex_q2_elim}, we see that $b_{3b}$ does not contribute to the flow of $\zeta_k$, while its contribution to the flow of $\nu_k$ and $\lambda_k$ is
\be
\begin{aligned}
k\partial_k\big(Z_k^\frac{3}{2} \nu_k\big) &=-C_4 +\dots,\\
k\partial_k\big(Z_k^\frac{3}{2} \lambda_k\big) &=\ \; C_4 +\dots,
\end{aligned}
\ee
where the dots stands for the contribution of other diagrams to the flow equation.
Therefore, after doing the rescalings in Eq.~\eqref{Eq:dimless_cpls}, one obtains the contribution of $b_{3b}$ to $b_\nu$ and $b_\lambda$ given in Eqs.~\eqref{Eq:bnu} and \eqref{Eq:blambda}, respectively,

\subsection{Comparison with the perturbative RG}

Let us briefly comment on the calculation of the diagrams using the perturbative RG.
The difference compared to the FRG case is that there is no explicit regulator present, and therefore, no derivative with respect to $k$ has to be taken and $\vec{q}^{\,2}_R$ is replaced by $\vec{q}^{\,2}$. 
In the case of the $b_{3b}$ diagram, calculated in the previous subsection in the FRG framework, one proceeds now from Eq.~\eqref{Eq:b3b_v2} using the result \eqref{Eq:w-int_b3b} for the frequency integral and setting $a_k$ to zero. 
Then, as discussed in Sec.~\ref{ss:mom-shell_RG}, one integrates the magnitude of $|\vec{q}|$ within the shell $[\Lambda/b,\Lambda]$ with $b=1+\rd b$ and expands in the external momentum.
Taking the $-\partial_{\rd b}$ derivative, one obtains the contribution to the $\beta$ functions in the form appearing in Eqs.~(15) and (16) of Ref.~\cite{caballero2018}.

With the momentum shell method sketched above, we have checked all $\beta$ functions calculated in Ref.~\cite{caballero2018}. 
We have found only one discrepancy, namely in the full expression of $\beta_u$, which was only given in Ref.~\cite{caballero_PhD}, since in Ref.~\cite{caballero2018}, the triangle contribution with four external legs was not considered.
Specifically, we disagree with the coefficient of the $\bar u\bar \nu^2$ term, which should be 3 rather than 9/2 in the expression for $C_2$ in Eq.~(4.6) of Ref.~\cite{caballero_PhD}.
The problematic contribution arises from diagram (j) in Fig.~4.1 of Ref.~\cite{caballero_PhD} (see also its Appendix C), which corresponds to the lower two diagrams in our Fig.~\ref{Fig:triangles}.

\bibliography{AMB+}

\end{document}